\documentclass[twocolumn,showpacs,preprintnumbers,superscriptaddress,amsmath,floatfix,amssymb,prd,nofootinbib]{revtex4}
\usepackage[colorlinks=true]{hyperref}
\usepackage{graphicx}
\usepackage{bm}
\newcommand{\comment}[1]{}

\newcommand{\lr}[1]{ \left( #1 \right) }
\newcommand{\lrs}[1]{ \left[ #1 \right] }

\newcommand{\vev}[1]{ \langle \, #1 \, \rangle }

\newcommand{\tr}{ {\rm Tr} \, }

\renewcommand{\Im}{ {\rm Im} \, }

\newcommand{\GeV}{{\rm GeV}}
\newcommand{\MeV}{{\rm MeV}}
\newcommand{\fm}{{\rm fm}}

\newcommand{\mymin}[1]{ {\rm min} \left( #1 \right) }
\newcommand{\mymax}[1]{ {\rm max} \left( #1 \right) }
\newcommand{\const}{ {\rm const}}
\renewcommand{\det}[1]{ {\rm det} \left( #1 \right) }

\newcommand{\sign}{ {\rm sign} \,  }

\newcommand{\expa}[1]{ \exp{\left( #1 \right)} }

\newcommand{\be}{\begin{equation}}
\newcommand{\ee}{\end{equation}}

\newcommand{\floor}[1]{\left\lfloor #1\right\rfloor}

\begin{document}
\sloppy

\title{Electric charge catalysis by magnetic fields and a nontrivial holonomy}

\author{F. Bruckmann}
\email{falk.bruckmann@ur.de}
\affiliation{Institut f\"{u}r Theoretische Physik, Universit\"{a}t Regensburg, D-93040 Regensburg, Germany}

\author{P. V. Buividovich}
\email{pavel.buividovich@ur.de}
\affiliation{Institut f\"{u}r Theoretische Physik, Universit\"{a}t Regensburg, D-93040 Regensburg, Germany}

\author{T. Sulejmanpasic}
\email{tin.sulejmanpasic@ur.de}
\affiliation{Institut f\"{u}r Theoretische Physik, Universit\"{a}t Regensburg, D-93040 Regensburg, Germany}

\begin{abstract}
 We describe a generic mechanism by which a system of Dirac fermions in thermal equilibrium acquires electric charge in an external magnetic field. To this end the fermions should have an additional quantum number, isospin or color, and should be subject to a second magnetic field, which distinguishes the isospin/color, as well as to a corresponding isospin chemical potential. The role of the latter can be also played by a nontrivial holonomy (Polyakov loop) along the Euclidean time direction. The charge is accumulated since the degeneracies of occupied lowest Landau levels for particles of positive isospin and anti-particles of negative isospin are different. We discuss two physical systems, where this phenomenon can be realized. One is monolayer graphene, where the isospin is associated with two valleys in the Brillouin zone and the strain-induced pseudo-magnetic field acts differently on charge carriers in different valleys. Another is hot QCD, for which the relevant non-Abelian field configurations with both nonzero chromo-magnetic field and a nontrivial Polyakov loop can be realized as calorons - topological solutions of Yang-Mills equations at finite temperature. The induced electric charge on the caloron field configuration is studied numerically. We argue that due to the fluctuations of holonomy external magnetic field should tend to suppress charge fluctuations in the quark-gluon plasma and estimate the importance of this effect for off-central heavy-ion collisions.
\end{abstract}

\date{July 1st, 2013}

\maketitle

\section{Introduction}
\label{sec:introduction}

 Interactions of chiral fermions with magnetic fields are the origin of a number of nontrivial physical phenomena, which have been intensively studied in the last few years \cite{MagneticQCDBook}. Some of the most notable examples of such phenomena are the Chiral Magnetic Effect (CME) \cite{Kharzeev:08:1, Kharzeev:08:2}, which is the generation of electric current in a system of chiral fermions at finite chiral chemical potential, and the Chiral Separation Effect (CSE) \cite{Son:04:2, Metlitsky:05:1}, the generation of chiral current in a system of chiral fermions at finite chemical potential. The most striking feature of such phenomena is that electric current is generated due to the magnetic field, so that no electric field is needed to initiate the transport of charge/chirality. Rather, this current has a topological nature related to the axial anomaly for chiral fermions and spin projection in the lowest Landau level \cite{Kharzeev:08:1, Sadofyev:11:1} and flows without dissipation even for interacting fermions \cite{Jensen:12:1, Banerjee:12:1, Metlitsky:05:1}. Experimentally, such phenomena are supposed to manifest themselves in specific features of the distributions of charged particles produced in heavy ion collisions \cite{Voloshin:04:1}.

 In this paper we describe a similar phenomenon, in which an electric charge, rather than current, is generated in a system of Dirac fermions in the presence of an external magnetic field. For this reason we call it ``charge catalysis by a magnetic field''. To this end the fermions, in addition to charge and chirality, should carry yet another quantum number, which we call isospin $\tau_3$ (not to be confused with the term isospin commonly used in particle physics) and assume that it takes values $\tau_3 = \pm 1$. In QCD, this additional quantum number is naturally associated with color. The fermions should also interact with an isospin magnetic field, or a chromomagnetic field in the case of QCD, for which the vector potential $A_{\mu}'$ enters the action as $i \bar{\psi} \tau_3 A_{\mu}' \gamma_{\mu} \psi$. We consider a grand canonical ensemble at nonzero isospin chemical potential $\mu_3$ which enters the action as $\mu_3 \bar{\psi} \tau_3 \gamma_{0} \psi$ and thus favors particles with $\tau_3 = +1$ and anti-particles with $\tau_3 = -1$ and disfavors particles with $\tau_3 = -1$ and anti-particles with $\tau_3 = +1$. For QCD, the role of imaginary isospin chemical potential can be also played by the time-like component of non-Abelian gauge field. The generated charge is also imaginary in this case, and shows up only in charge fluctuations after the functional integration over the non-Abelian gauge field. While the mechanism of the described effect is not directly related to the axial anomaly, it is still similar to CME in the sense that charge generation (which also manifests itself as spatial separation of positive and negative charges, as we will demonstrate in Subsection \ref{subsec:charge_halo}) is caused by magnetic, rather than electric field, and has a purely quantum origin related to the two-dimensional index theorem. As we argue in Subsection \ref{subsec:temperature}, for a classical gas of charged particles this effect would be absent.

 It should be stressed that it is the assumption of a grand canonical ensemble that makes the generation of charge possible: the system is not isolated, but rather coupled to an environment, with which it can exchange particles. Thus the extra charge comes from the environment, and charge conservation is not violated.  The system then acquires electric charge \emph{only} if all three factors are present: an ordinary magnetic field, an isospin magnetic field and a nonzero isospin chemical potential $\mu_3$. At vanishing temperature, the latter can be arbitrarily small. The mechanism of this phenomenon can be also reversed, so that an isospin polarization is generated at nonzero ordinary chemical potential $\mu$ (which couples to electric charge), again in the presence of both ordinary and isospin magnetic fields.

 We should note here that this work was in fact originally motivated by an attempt to describe the Chiral Magnetic Effect \cite{Kharzeev:08:1, Kharzeev:08:2} by using calorons, topologically nontrivial classical solutions of finite-temperature Yang-Mills equations \cite{Harrington:78:1,Kraan:98:2, Lee:98:1}, in external magnetic fields as a model for topological transitions in QCD at finite temperature (see Subsection \ref{subsec:caloron}). While we have found no traces of electric current flowing along the magnetic field, we have noticed the emergence of a nonzero imaginary charge density induced by a magnetic field parallel to the caloron axis. The present work was initiated as an attempt to understand this peculiar effect.

 The paper is organized in the following way: in Section \ref{sec:uniform} we start with a simple description of charge catalysis at zero temperature and at small values of the isospin chemical potential, and in Subsection \ref{subsec:temperature} discuss the case of arbitrary temperatures/chemical potentials. The dependence of the induced charge on isospin chemical potential turns out to be quite nontrivial and is in some sense similar to Shubnikov-de Haas oscillations. In Subsection \ref{subsec:charge_halo} we consider the case of spatially localized magnetic and isospin magnetic fields. We demonstrate that even in the case when the net induced charge is zero, there is still an excess of charge in the region with nonzero field strength, which is compensated by an excess of charge of opposite sign in a ``halo'' around that region. Thus in such a setting the magnetic field induces spatial charge separation. In Section \ref{sec:graphene} we illustrate the described mechanism on a very simple and at the same time physically relevant example of strained monolayer graphene. In this case the isospin index corresponds to two distinct valleys in the graphene Brillouin zone and the isospin magnetic field is the pseudo-magnetic field arising due to mechanical strain \cite{KatsnelsonGraphene, Vozmediano:10:1}. In Section \ref{sec:qcd} we discuss possible manifestations of charge catalysis in finite-temperature QCD, for which the role of isospin is played by the color of quarks and that of the isospin chemical potential by a nontrivial holonomy (Polyakov loop) of a gauge field configuration. We first consider a simple example of a constant non-Abelian gauge field and demonstrate that while the net electric charge vanishes in this case due to integration over all values of the holonomy, the described phenomenon still manifests itself in the suppression of charge fluctuations. In Subsection \ref{subsec:caloron} we consider a more realistic example of an $SU\lr{2}$ caloron gauge field configuration. We show numerically that the induced charge is localized between the monopole and the anti-monopole which constitute the caloron, that is, in the region where the chromo-magnetic field is approximately constant. Finally, in Subsection \ref{subsec:experiment} we roughly estimate the possible contribution of the discussed effect to charge fluctuations in off-central heavy-ion collisions, using both the model of a uniform chromomagnetic field with fluctuating holonomy (Subsection \ref{subsec:const_field_qcd}) as well as a model of a dilute caloron gas \cite{Ilgenfritz:07:1}.

\section{Charge catalysis in uniform magnetic fields and isospin chemical potentials: mechanism}
\label{sec:uniform}

\subsection{The limit of zero temperature and small isospin chemical potential}
\label{subsec:zero_T}

 Let us denote the strength of the ordinary magnetic field as $B$, and the strength of the isospin magnetic field as $F$. For simplicity we first assume that the fermions are $\lr{2 + 1}$-dimensional, but our arguments can be easily generalized to $\lr{3 + 1}$-dimensional case. The magnetic and the isospin magnetic fields in $\lr{2+1}$ dimensions have only a single component, $B \equiv B_{xy} = \partial_x A_y - \partial_y A_x = \const$ and $F \equiv F_{xy} = \partial_x A_y' - \partial_y A_x' = \const$. The energy spectrum of fermions in such a background is given by the Landau levels (we assume magnetic fields in the CGS units\footnote{For SI units one would take away the factor of $c$ in the denominator.})
\begin{equation}
\label{landau_levels}
 E_{n}^{\lr{q, \tau_3}} = \sqrt{\frac{2 \hbar e v_F^2}{c} |q B + \tau_3 q F| n}\, ,
\end{equation}
where $n = 0, 1, 2, \ldots$, the labels $q = \pm 1$ and $\tau_3 = \pm 1$  denote particle/antiparticle states and the isospin quantum number, respectively, $e$ is the electron charge, $c$ is the speed of light and $v_F$ is the Fermi velocity. For relativistic fermions (electrons or quarks in quark-gluon plasma) $v_F = c$, but for Dirac quasiparticles in graphene $v_F$ is significantly smaller, $v_F \approx c/300$ \cite{KatsnelsonGraphene} \footnote{For graphene applications, depending on the definition of the strain magnetic field, the factor of $c/v_F \approx 300$ may appear in front of $F$, while for QCD, where $F$ is the chromomagnetic field, a factor of $g/e$ may appear.}. Finally, for QCD applications, the electron charge $e$ in (\ref{landau_levels}) should be replaced by the quark charge $-1/3 e$ or $2/3 e$.

The degeneracies of the Landau levels are
\begin{equation}
 g_n |q \Phi_B + \tau_3 q \Phi_F|
\end{equation}
where $g_0 = 1$, $g_{n > 0} = 2$ and $\Phi_B = \lr{2 \pi \hbar}^{-1}\int\limits_{S} dx \, dy \, e B$ and $\Phi_F = \lr{2 \pi \hbar}^{-1}\int\limits_{S} dx \, dy \, e F$ are the fluxes of the magnetic and the isospin magnetic fields through the area $S$ of the system. For definiteness, we have assumed periodic boundary conditions for spatial directions, which implies that $\Phi_B$ and $\Phi_F$ take integer values \cite{Wiese:08:1}. From now on we work in natural units with $\hbar = e = c = 1$ in order to shorten the notation (in particular, this means that the physical electron or quark charge is absorbed into the units in which $F$ and $B$ are expressed). It also follows from (\ref{landau_levels}) that the effect of the Fermi velocity $v_F$ is to simply rescale all the energies by $v_F$. Thus for the sake of simplicity we also assume that $v_F = c = 1$. The results of Section \ref{sec:graphene}, in which we consider charge catalysis in graphene, will not depend on $v_F$ anyway.

\begin{figure}[h]
  \centering
  \includegraphics[width=7.5cm]{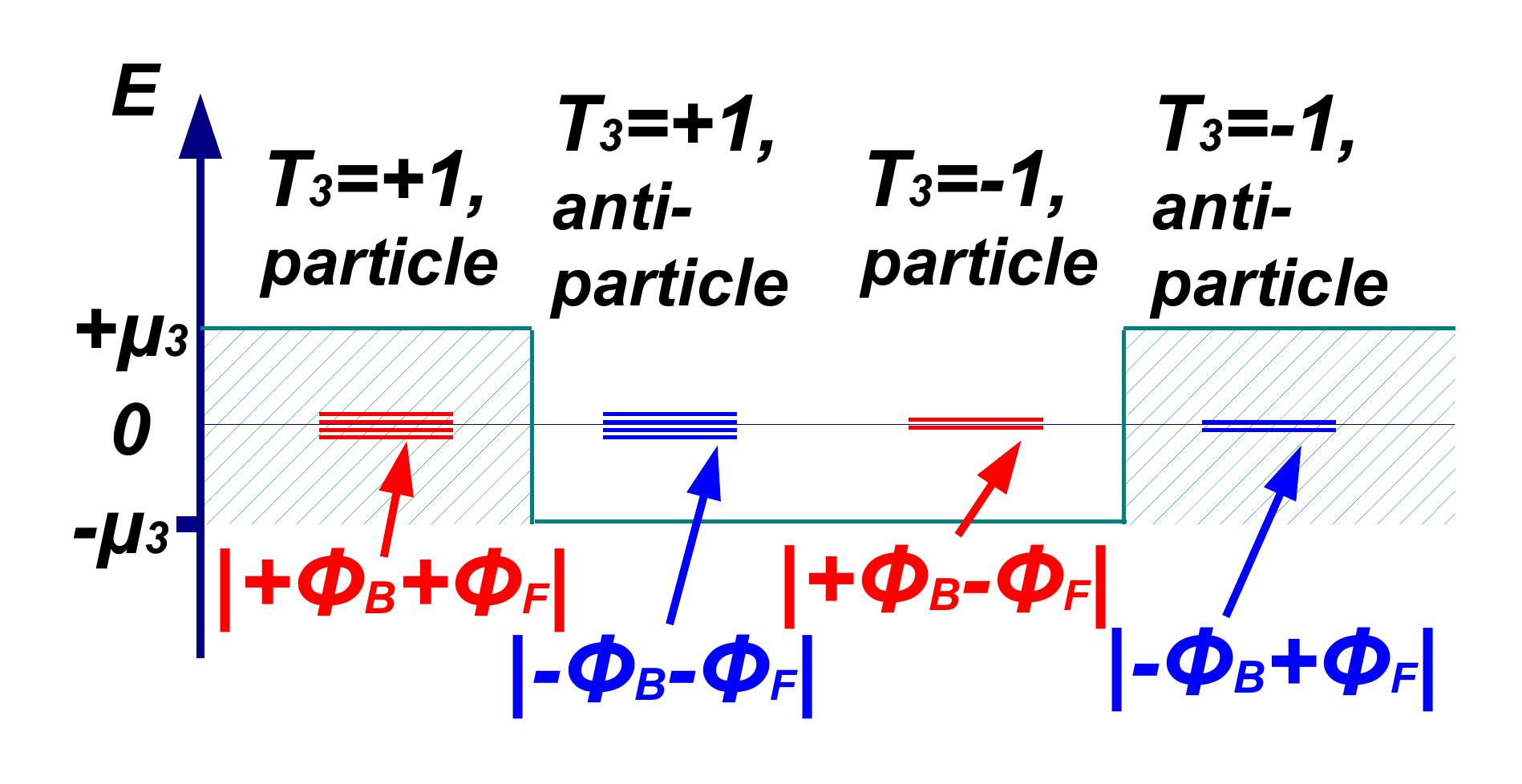}\\
  \caption{Schematic illustration of the degeneracies of the lowest Landau levels and the Fermi energies for particles and anti-particles with $\tau_3 = \pm 1$ for $\Phi_F = 3$, $\Phi_B = 1$. The number of short horizontal lines in each column denotes the degeneracy of the lowest Landau level for the corresponding quantum numbers and the long horizontal line denotes the zero energy level.}
  \label{fig:levels_shift_general}
\end{figure}

 At nonzero isospin chemical potential $\mu_3$ the Fermi energies for particles/anti-particles with $\tau_3 = \pm 1$ become $E_F^{\lr{q, \tau_3}} = q \, \tau_3 \, \mu_3$. For simplicity let us first assume that $\mu_3 < \mymin{\sqrt{2|F-B|}, \sqrt{2|F+B|} }$ and that the temperature is zero (these conditions will be relaxed in the next subsection). In this case only the lowest Landau levels for particles ($q = +1$) with $\tau_3 = +1$ and anti-particles ($q = -1$) with $\tau_3 = -1$ will be occupied. The degeneracies of these levels are $|\Phi_B + \Phi_F|$ and $|-\Phi_B + \Phi_F|$, respectively, and thus different (see Fig.~\ref{fig:levels_shift_general}). Therefore, if both $\Phi_B$ and $\Phi_F$ are present, there is an unequal number of particles and antiparticles on these levels, and the system has a nonzero electric charge:
\begin{align}
\label{charge_generated}
 Q =&\, |\Phi_B + \Phi_F| - |-\Phi_B + \Phi_F|
 \nonumber \\
 = &\,
 2 \, \sign(\Phi_F \, \Phi_B) \, \mymin{|\Phi_B|, |\Phi_F|} .
\end{align}
If the strength of the isospin magnetic field $F$ is kept constant, the charge grows linearly with $B$ for $|B| < |F|$ and then saturates at $Q = \pm \, 2 |\Phi_F|$. If $|B| < |F|$, increasing the flux $\Phi_B$ by one flux quantum produces two elementary charges. We also note that since the existence of zero energy level in the magnetic field is topologically protected in $\lr{2+1}$ dimensions, the result (\ref{charge_generated}) is also valid for non-uniform gauge fields as long as $\mu_3$ is smaller than the first non-zero energy level.

 Finally, we note that the mechanism described above can be reversed by switching on the ordinary chemical potential which couples to electric charge and thus increases the Fermi energy for particles and decreases it for anti-particles. Then only the lowest Landau levels for particles 
 will be occupied, independently of $\tau_3$. The degeneracies of the lowest Landau levels for particles with $\tau_3 = +1$ and $\tau_3 = -1$ are again given by $|\Phi_B + \Phi_F|$ and $|-\Phi_B + \Phi_F|$, respectively, and thus different again. Therefore, a nonzero isospin charge $Q_3 = |\Phi_B + \Phi_F| - |-\Phi_B + \Phi_F|$ is produced.

\subsection{Charge catalysis at arbitrary temperatures and isospin chemical potentials}
\label{subsec:temperature}

\begin{figure}[b]
   \centering
   \includegraphics[width=9cm]{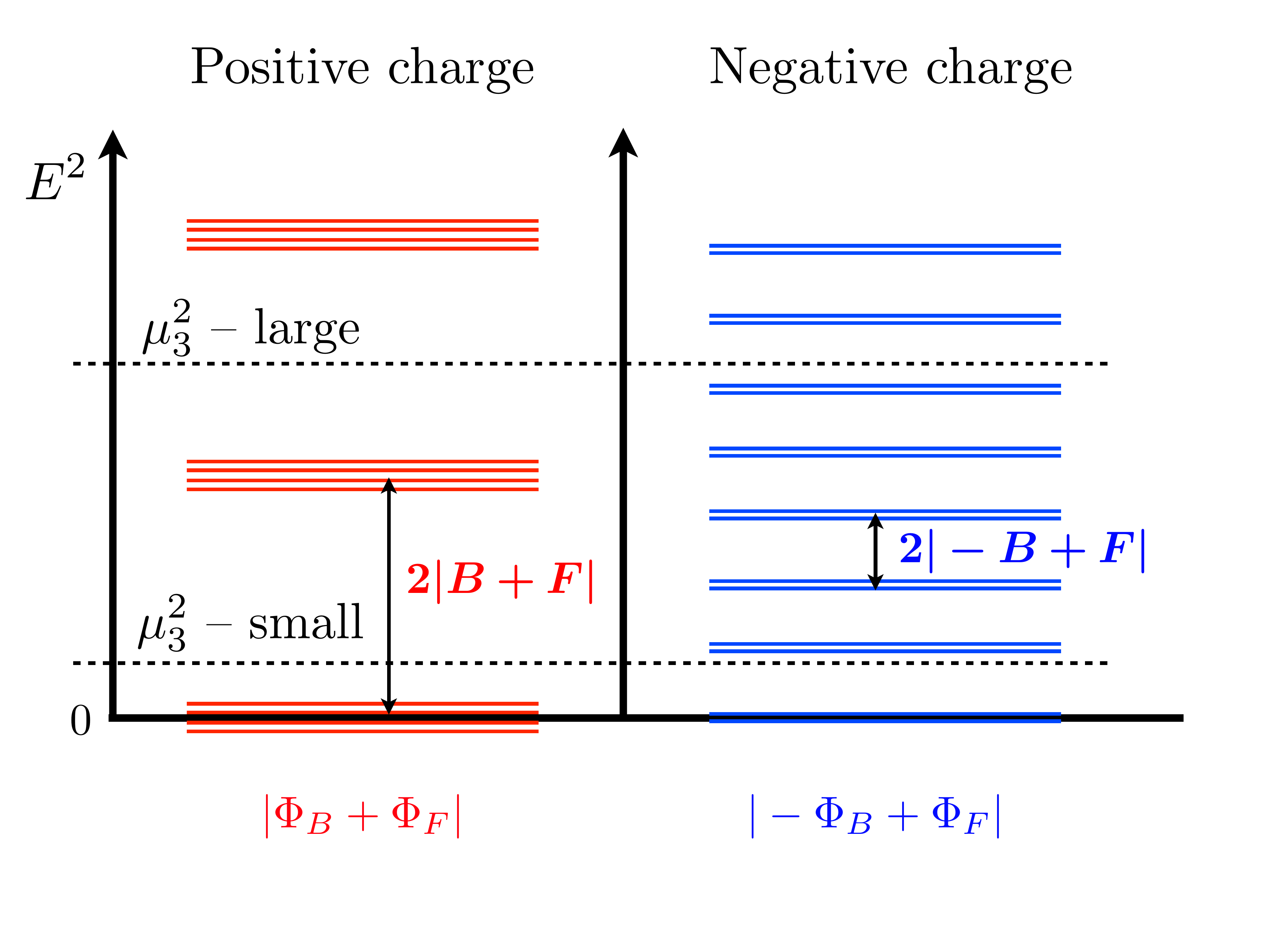}
   \caption{Occupied Landau levels for particles in the $\tau_3 = +1$ sector (leftmost column on Fig. \ref{fig:levels_shift_general}) and anti-particles in the $\tau_3 = - 1$ sector (rightmost column on Fig. \ref{fig:levels_shift_general}) at some finite value of an isospin chemical potential. The level degeneracies are labeled below as $|\pm \Phi_B + \pm \Phi_F|$. Horizontal dotted lines illustrate two values of $\mu_3$, one ``small'' (i.e. $\mu_3^2<2|-B+F|$) and one large.}
   \label{fig:high_mu_sketch}
\end{figure}

 Let us now discuss how the generated charge (\ref{charge_generated}) is changed at higher values of isospin chemical potential and/or at nonzero temperature, when higher Landau levels are also populated. Consider first the case of zero temperature but an arbitrary value of isospin chemical potential. The equation (\ref{charge_generated}) is then easily generalized to
\begin{align}
\label{charge_generated_high_mu}
 Q = &\,|\Phi_B + \Phi_F| - |-\Phi_B + \Phi_F|
 \nonumber\\
 &\,+
 2\floor{ \frac{\mu_3^2}{2|B+F|} } \, |\Phi_B+\Phi_F|
 \nonumber\\
 &\,-
 2\floor{ \frac{\mu_3^2}{2|-B+F|} } \, |-\Phi_B+\Phi_F|  ,
\end{align}
where $\floor{\ldots}$ is the floor function.  The first line coming from the lowest Landau levels is the same as in Eq.~\eqref{charge_generated}. The second line in (\ref{charge_generated_high_mu}) counts the charge of the particles on higher Landau levels (left side of Fig.~\ref{fig:high_mu_sketch} and the leftmost column in Fig.~\ref{fig:levels_shift_general}), while the last line comes from the contribution of anti-particles (right side of Fig.~\ref{fig:high_mu_sketch} and the rightmost column in Fig.~\ref{fig:levels_shift_general}). The factor of two in the second and the third lines comes from the spin degeneracy $g_{n>0}=2$ of the higher Landau levels.

\begin{figure}[htbp]
   \centering
   \includegraphics[width=.5\textwidth]{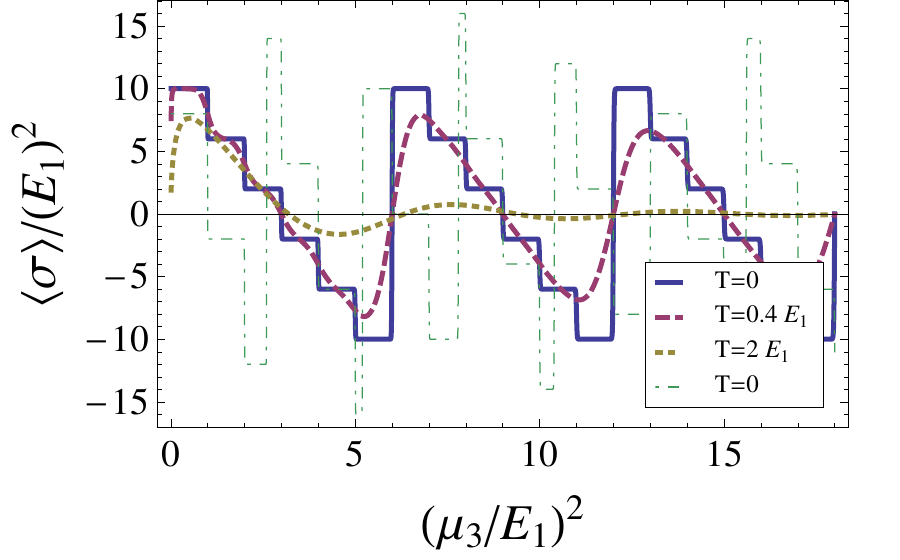}
   \includegraphics[width=.5\textwidth]{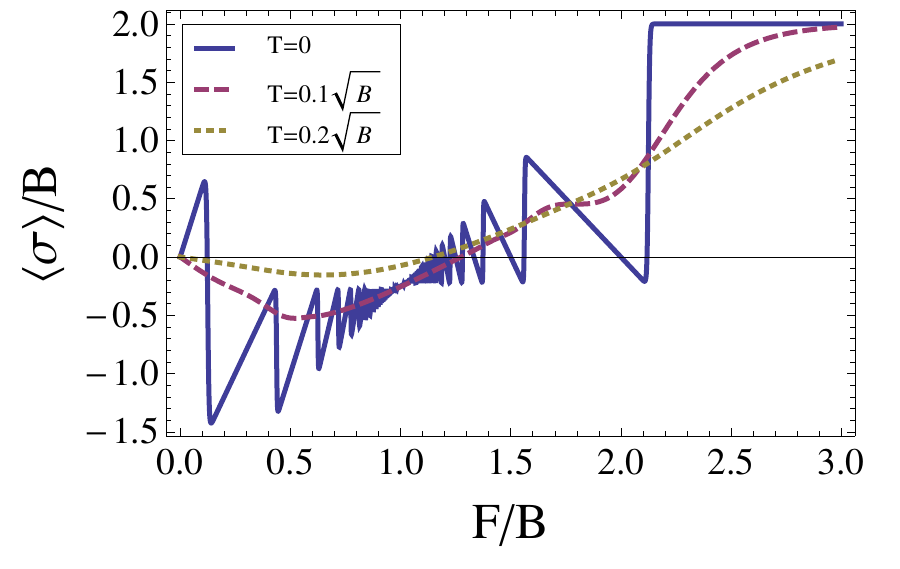}
   \caption{Top: the density $\vev{\sigma} = Q/S$ of generated charge as a function of isospin chemical potential squared (in units of the next-to-lowest Landau level $E_1^2= 2|B-F|$) for $(F-B)/(F+B)=1/6$ (blue solid line, red dashed line and yellow dotted line) and $(F-B)/(F+B)=5/13$ (green dot-dashed line) and different temperatures. Bottom: the density of generated charge at $\mu_3 = 1.5 \, \sqrt{B}$ as a function of isospin magnetic field $F$ for different temperatures.}
   \label{fig:charge_generated_high_mu}
\end{figure}

 The expression (\ref{charge_generated_high_mu}), although being conceptually very simple, leads to quite a nontrivial dependence of the generated charge on the chemical potential and the magnetic field, see Fig.~\ref{fig:charge_generated_high_mu} for an illustration. Depending on how many Landau levels occupied by particles or anti-particles lie below $\mu_3$, the charge $Q$ can take both positive and negative values.
The charge decreases if one of the more closely spaced negatively charged states sinks below the Fermi energy and increases if this happens for one of the more widely separated positively charged states. It is also easy to prove that the charge (\ref{charge_generated_high_mu}) is a periodic function of $\mu_3^2$ with a period equal to $\Delta \mu_3^2 = \frac{4 \pi}{S} \, {\rm LCM}\lr{|\Phi_F + \Phi_B|, |\Phi_F - \Phi_B|}$, where ${\rm LCM}\lr{n, m}$ stands for the least common multiple of two integers $n$ and $m$.

 The dependence of the induced charge density on the magnetic field at fixed isospin chemical potential exhibits oscillations similar to Shubnikov-de Haas oscillations, see Fig.~\ref{fig:charge_generated_high_mu} below. However, while in the latter case the period of oscillations tends to zero at zero magnetic field, here this happens at $F$ close to $B$, when the spacing between energy levels for anti-particles with $\tau_3 = -1$ becomes very small.

 In case of finite temperature the floor functions in (\ref{charge_generated_high_mu}) are smoothed out by the corresponding Fermi factors, so that the charge is given by
\begin{align}
\label{charge_generated_finite_t}
 Q = \sum\limits_{n = 0}^{+\infty}g_n
 \Big(&\,
 \frac{|\Phi_B + \Phi_F|}{1 + e^{\frac{E_n^{\lr{++}} - \mu_3}{T}}}
 -
 \frac{|\Phi_B + \Phi_F|}{1 + e^{\frac{E_n^{\lr{--}} + \mu_3}{T}}}
 \nonumber\\
 +&\,
 \frac{|\Phi_B - \Phi_F|}{1 + e^{\frac{E_n^{\lr{+-}} - \mu_3}{T}}}
 -
 \frac{|\Phi_B - \Phi_F|}{1 + e^{\frac{E_n^{\lr{-+}} + \mu_3}{T}}}\,\Big) \,.
\end{align}
We have evaluated these sums numerically. The resulting dependence of the generated charge on $\mu_3$, $F$ and $B$ is illustrated in Fig.~\ref{fig:charge_generated_high_mu}. One can see that now the dependence of the generated charge on $\mu_3$, $F$ and $B$ becomes continuous. However, the charge still changes sign as $\mu_3$ is changed. Since the asymptotic density of states for Landau levels in both sectors with $\tau_3 = \pm 1$ is equal (since smaller spacing between Landau levels is compensated for by smaller level degeneracy), at very high temperatures and/or isospin chemical potentials the generated charge should vanish. Asymptotic equality of the density of states in both sectors implies also that the described effect should vanish in the classical limit, when a macroscopically large number of levels is occupied.

\subsection{Spatial charge separation}
\label{subsec:charge_halo}

 As discussed in the Introduction, the generation of charge is only possible for the grand canonical ensemble, that is, for a system coupled to some environment, from which the charge is ``borrowed'', and charge conservation is not violated. In this Subsection we investigate in more details the situation in which the charge density is generated locally, but the net charge is still zero. We show that this situation can be interpreted as spatial charge separation.

 Let us formally assume that the space is infinite and the magnetic field strengths are described by some localized functions\footnote{This means that they decay sufficiently quickly at infinity.}. In this case normalizable eigenmodes with zero energy exist also for non-integer values of the fluxes $\Phi_B$, $\Phi_F$ \cite{Aharonov:78:1}. Their number is given by the floor functions of $\Phi_B$ and $\Phi_F$, so that the equation (\ref{charge_generated}) reads in this case
\begin{eqnarray}
\label{charge_generated_aharonov}
 Q = \floor{|\Phi_F+\Phi_B|} - \floor{|-\Phi_F+\Phi_B| } .
\end{eqnarray}
In deriving (\ref{charge_generated_aharonov}) we have again assumed that there is a small but finite isospin chemical potential $\mu_3$ so that only the particle zero energy states with $\tau_3=+1$ and antiparticle zero energy states with $\tau_3=-1$ are filled.

 We now restrict ourselves to the situation in which both fluxes $\Phi_F$ and $\Phi_B$ are nonzero, but the total charge (\ref{charge_generated_aharonov}) is still zero. For example, one can take $\Phi_F = n + 1/2$ and $0 < \Phi_B < 1/2$. If the region in which the magnetic field is localized is sufficiently small, one can still have very large magnetic fields locally. We would like to study here whether the charge is generated locally in this case, and how it is distributed.

 As the simplest example let us assume that both fields $B$ and $F$ take constant values in the region $r \le R$, where $r$ is the radial coordinate in two spatial dimensions, and are equal to zero outside of this region. The wave functions of the zero energy states can be found analytically as \cite{Aharonov:78:1}
\begin{eqnarray}
\label{eq:2dzeromodes}
 \psi_n\lr{x,y} = c_n^{\pm} \, z^{n} \, e^{-\phi^{\pm}\lr{x, y}}s^{\uparrow},
 \quad n = 0 , 1, \ldots, N^{\pm} ,
\end{eqnarray}
where $z = x + iy$, $s^+$ is the unit spinor which is the eigenstate of the projection of the spin operator on the magnetic field with eigenvalue $s = +1/2$, $N^{\pm} = \floor{|\Phi_F \pm \Phi_B|} - 1$ and $\phi\lr{x, y}$ is the solution of the Poisson equation
\begin{eqnarray}
\label{eq:poisson}
 \lr{\partial_x^2 + \partial_y^2} \phi^{\pm}\lr{x, y} = B\lr{x, y} \pm F\lr{x, y} .
\end{eqnarray}
It is easy to see that the eigenvectors (\ref{eq:2dzeromodes}) are orthogonal to each other for cylindrically symmetric distributions of magnetic field (in the absence of cylindrical symmetry this might not be the case). Solving the equation (\ref{eq:poisson}) for magnetic field which is localized in a cylinder of radius $R$ we obtain
\begin{eqnarray}
\label{poisson_solution}
\phi^{\pm}\lr{r} =
 \begin{cases}
  \lr{B \pm F} \, \lr{r^2 - R^2}/4 , &  r \le R  \\
  \lr{B \pm F} \, R^2/2 \, \ln\lr{r/R},  & r > R \\
 \end{cases}
\end{eqnarray}
The wave functions (\ref{eq:2dzeromodes}) are then:
\begin{eqnarray}
\label{eq:halo_states}
 \psi^{\pm}_n\lr{x, y} = c_n^{\pm} z^{n} s^\uparrow
 \begin{cases}
  e^{\frac{1}{4}|F \pm B|\lr{R^2 - r^2}} & r \le R\\
  \lr{\frac{R}{r}}^{\frac{1}{2}{|F \pm B| R^2}} & r > R \\
\end{cases}.
\end{eqnarray}
Although the normalization constants $c_n^{\pm}$ can be calculated analytically, the resulting expressions are not very instructive and we do not give them here. It is important to note, however, that for large $|\Phi_B \pm \Phi_F|$ and $n \ll |\Phi_B \pm \Phi_F|$ they behave as
\begin{eqnarray}
\label{eq:halos_norm_lim}
 c_n^{\pm} \sim \frac{1}{\sqrt{\pi}}\lr{\frac{|F \pm B|}{2}}^{\frac{n+1}{2}} \frac{e^{-|\Phi_B \pm \Phi_F|}}{\sqrt{n!}}
\end{eqnarray}

\begin{figure}[htbp]
   \centering
   \includegraphics[width=8.5cm]{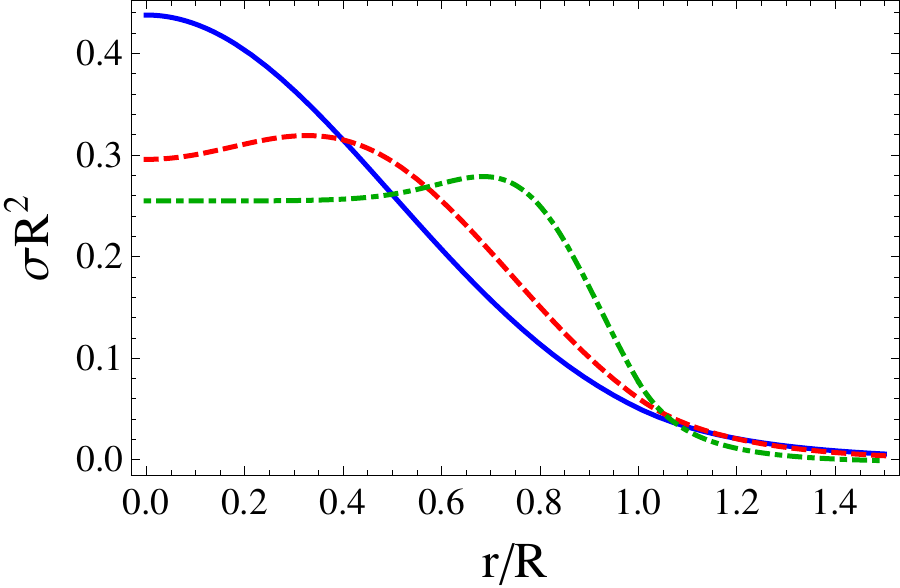}\\
   \includegraphics[width=8.5cm]{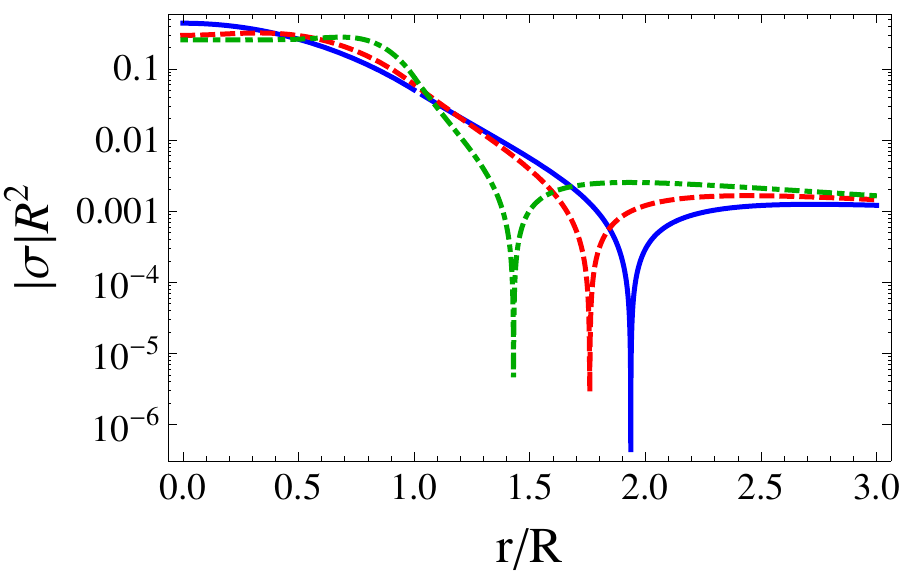}\\
   \caption{Top: Charge density (\ref{eq:halo_ch_dens}) as a function of the polar coordinate $r$. The fluxes are $\Phi_B = 0.4 $ and $\Phi_F = 1.5, 2.5, 10.5$  for solid-blue, dashed-red and dot-dashed green lines, respectively. Bottom: The absolute value of same charge density on a logarithmic plot, to emphasize the region with negative charge density at $r \gtrsim R$.}
 \label{fig:charge_halo}
\end{figure}

 Under the assumption that all the zero energy states are occupied, the charge density is given by
\begin{eqnarray}
\label{eq:halo_ch_dens}
 \sigma\lr{x, y} = \sum\limits_{n=0}^{N^+} |\psi_n^+\lr{x, y}|^2 - \sum\limits_{n=0}^{N^-} |\psi_n^-\lr{x, y}|^2  ,
\end{eqnarray}
where $N^{\pm} = \floor{|\Phi_B \pm \Phi_F|} - 1$. This result is illustrated on Fig.~\ref{fig:charge_halo}, where we plot the charge density as a function of the radial coordinate $r$ for the values of $\Phi_B$ and $\Phi_F$ for which the net generated charge (\ref{charge_generated_aharonov}) vanishes. One can see that while positive charge is accumulated in the region with nonzero field strength ($r < R$), the excess of negative charge is accumulated in the exterior of that region. At large distances $r \gg R$, the negative charge density behaves as
\begin{align}
\label{charge_dens_asymptotics}
 \sigma\lr{r} \sim &\, 1/r^{2 + 2 \epsilon},
 \nonumber \\
 \epsilon = &\,\mymin{ |\Phi_B \pm \Phi_F| - \floor{|\Phi_B \pm \Phi_F|}} ,
\end{align}
where the last expression is just the smallest fractional part of the fluxes $|\Phi_B \pm \Phi_F|$. Thus the charge distribution has a negative ``heavy tail'' which compensates for the positive charge accumulated in the region with nonzero field strengths. Such exact compensation becomes obvious if one integrates (\ref{eq:halo_ch_dens}) over the whole space, taking into account the normalization of the eigenmodes $\psi_n^{\pm}\lr{x, y}$: the total charge is then just $Q = N_+ - N_- = \floor{|\Phi_F+\Phi_B|} - \floor{|-\Phi_F+\Phi_B| }$, which is equal to zero for field configurations under consideration ($\Phi_F = n + 1/2$, $0 < \Phi_B < 1/2$). In order to better illustrate the existence of this heavy tail, on the lower plot of Fig.~\ref{fig:charge_halo} we show the absolute value of the charge density in a logarithmic scale over a wider range of $r$.

 Let us now consider charge density for $r \ll R$ and assume that the fluxes $|\Phi_B \pm \Phi_F|$ are large. Since at small $r$ the eigenfunctions (\ref{eq:halo_states}) behave as $\psi^{\pm}_n\lr{r} \sim r^n$, the contributions of states with $n \gg 1$ are strongly suppressed, and one can safely extend the summation over $n$ in (\ref{eq:halo_ch_dens}) to infinity. Using now the asymptotic expression (\ref{eq:halos_norm_lim}) for the normalization constants $c_n^{\pm}$, we find for the charge density
\begin{align}
\label{charge_dens_small_r}
 \sigma\lr{r\ll R} \approx &\,
 \sum\limits_{n=0}^{+\infty} \frac{1}{\pi}\frac{|B+F|^{n+1} r^{2n}}{2^{n+1} \, n!} \, e^{-|B+F|r^2/2}
 \nonumber \\ -&\,
 \sum\limits_{n=0}^{+\infty} \frac{1}{\pi}\frac{|B-F|^{n+1} r^{2n}}{2^{n+1} \, n!} \, e^{-|B-F|r^2/2}
\nonumber \\
\approx&\,
\frac{1}{2 \pi} \, \lr{|B + F| - |B - F| } , \
\end{align}
which is is precisely the charge density which can be obtained from the expression (\ref{charge_generated}) valid in the case of uniform fields. We thus conclude that even if the system does not exchange electric charge with the environment, the charge generation still takes place due to spatial charge separation, and for sufficiently large fluxes $|\Phi_B \pm \Phi_F|$ the charge density in the center of a region with nonzero field strength is given precisely by (\ref{charge_generated}). One can see from Fig.~\ref{fig:charge_halo} that the charge density near $r = 0$ is indeed very close to this asymptotic value (which is equal to $\sigma \, R^2 = 2 \Phi_B/\pi$ if $\Phi_B < \Phi_F$) for $\Phi_F \gtrsim 2.5$. For smaller values of $\Phi_F$ the charge density is somewhat higher near the origin and smaller at the periphery. The requirement of sufficiently large flux simply means that the typical extent of the wave functions of the Landau states $R_B \sim 1/\sqrt{B}$ should be much smaller than $R$, so that the fermions effectively do not feel the boundary of the region with nonzero $F$ and $B$.

\begin{figure}[htbp]
   \centering
   \includegraphics[width=0.5\textwidth]{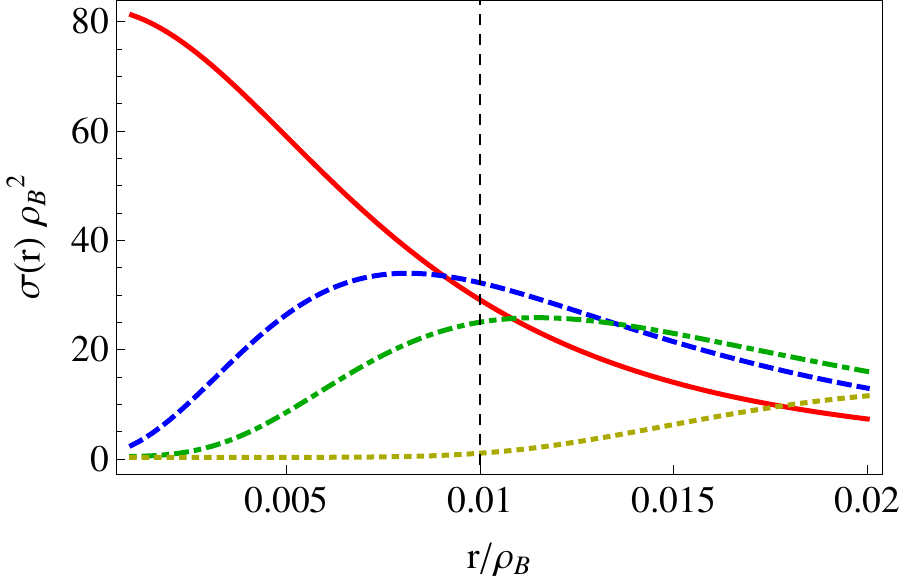}\\
   \includegraphics[width=0.5\textwidth]{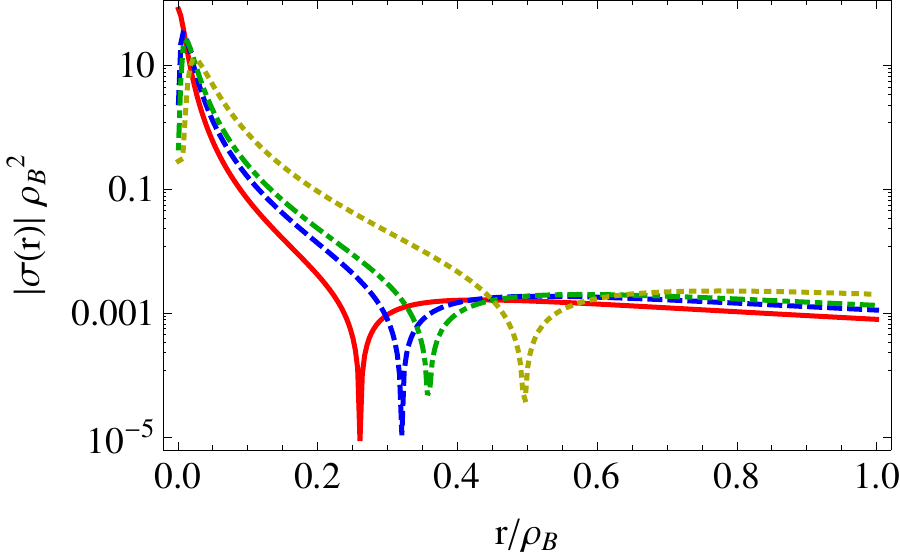}\\
   \caption{Top: Charge density (\ref{charge_dens_smooth}) in units of $\rho_B^2$ as a function of $r$ for $\Phi_B = 0.4$, $\rho_F/\rho_B = 0.01$ and $\Phi_F = 1.5$ (solid red line), $\Phi_F = 2.5$ (dashed blue line) $\Phi_F=3.5$ (dot-dashed green line) and $\Phi_F = 10.5$ (dotted yellow line). The vertical thin dashed line indicates the value of $\rho_F$. Bottom: The absolute value of the same charge densities on a logarithmic scale and for a much larger range of $r$.}
   \label{fig:smoothfields}
\end{figure}

Another interesting example is that of magnetic fields with the following profile
\begin{eqnarray}
\label{smooth_field_profiles}
 B\lr{r} = \frac{2 \Phi_B \, \rho_B^2}{\lr{r^2 + \rho_B^2}^2},
 \quad
 F\lr{r} = \frac{2 \Phi_F \, \rho_F^2}{\lr{r^2 + \rho_F^2}^2} ,
\end{eqnarray}
for which the zero energy eigenstates again take the form (\ref{eq:halo_states}) with the functions $\phi^{\pm}\lr{x, y}$ given by
\begin{eqnarray}
\label{smooth_field_potential}
 \phi^{\pm}\lr{r} = \Phi_B/2 \, \ln\lr{1 + r^2/\rho_B^2}
 \nonumber \\ \pm\,
 \Phi_F/2 \, \ln\lr{1 + r^2/\rho_F^2} .
\end{eqnarray}
Note that for unit flux this profile is the profile of the winding number density of the soliton in the two-dimensional $O\lr{3}$ model, where this density can also be interpreted as the magnetic field of an auxiliary Abelian gauge field \cite{RajaramanSolitonsInstantons}.

The charge density in this case reads
\begin{eqnarray}
\label{charge_dens_smooth}
 \sigma\lr{r} = \sum\limits_{n=0}^{N^+}|c^+_n|^2 r^{2n} \,
 \lr{1+\frac{r^2}{\rho_{F}^2}}^{-\Phi_F}
 \lr{1+\frac{r^2}{\rho_{B}^2}}^{-\Phi_B}
 \nonumber \\ -
 \sum\limits_{n=0}^{N^-}|c^-_n|^2 r^{2n} \,
 \lr{1+\frac{r^2}{\rho_{F}^2}}^{-\Phi_F}
 \lr{1+\frac{r^2}{\rho_{B}^2}}^{+\Phi_B} ,
\end{eqnarray}
where we have assumed that $\Phi_F>\Phi_B$. In Fig.~\ref{fig:smoothfields} we plot the charge density (\ref{charge_dens_smooth}) for the case of a strong $F$ field localized in a small region with $\rho_F \ll \rho_B$. It is interesting to note that in this case the charge density has a dip in the center (although remaining positive for $r \lesssim \rho_F$). This dip emerges due to the suppression of the densities of eigenstates with large $n$ by a factor proportional to $r^{2 n}$. In the case of uniform fields this contribution conspired with the gaussian envelope to make the distribution uniform. Here, however, the envelope is not gaussian and the effects of modes with $n > 0$ is to create a dip in the center of the region with nonzero strength of isospin magnetic field $F$. The mode with $n = 0$, which is monotonically decreasing with $r$, still creates a finite charge density at $r = 0$, but this contribution becomes smaller as compared to the contribution of higher modes at intermediate values of $r$. Note also that if $\mymax{N_+, N_-} \le 1$, only the $n=0$ mode contributes, and the dip does not exist (see, for example, the solid red curve on Fig. \ref{fig:smoothfields}, for which $\Phi_B = 0.4$, $\Phi_F = 1.5$). Again, since $ \floor{|\Phi_F+\Phi_B|} - \floor{|-\Phi_F+\Phi_B| }=0$, the excess of positive charge around $r = 0$ must be compensated by a negative charge density at large $r$, which decays as (\ref{charge_dens_asymptotics}). In order to illustrate this slow decay of negative charge density, on the second plot on Fig.~\ref{fig:smoothfields} we plot the absolute value of $\sigma\lr{r}$ at large $r \gg \rho_F$ on a logarithmic scale. One can see that indeed at some large $r \gg \rho_F$ the charge density changes sign and then decays very slowly.

\section{Charge catalysis in strained graphene}
\label{sec:graphene}

 Let us demonstrate now how the described phenomenon can be realized in graphene, a two-dimensional hexagonal crystal lattice of carbon atoms. The dispersion relation of valence electrons in graphene has two zeros - the Dirac points $K^{+}$ and $K^{-}$. The vicinities of these points are called valleys $K^{+}$ and $K^{-}$, respectively. The low-energy excitations in both valleys can be described as Dirac fermions \cite{KatsnelsonGraphene}, which couple to magnetic fields in the usual way. However, by applying mechanical strain in certain directions one can also induce a pseudo-magnetic field which has opposite signs for the two valleys, but does not distinguish between particles and holes (playing the role of anti-particles) \cite{Vozmediano:10:1}. This pseudo-magnetic field has been recently studied experimentally, and Landau levels which correspond to pseudo-magnetic field strengths larger than $300$ Tesla have been observed \cite{Crommie:10:1}.

\begin{figure}[htpb]
  \centering
  \includegraphics[width=6.5cm]{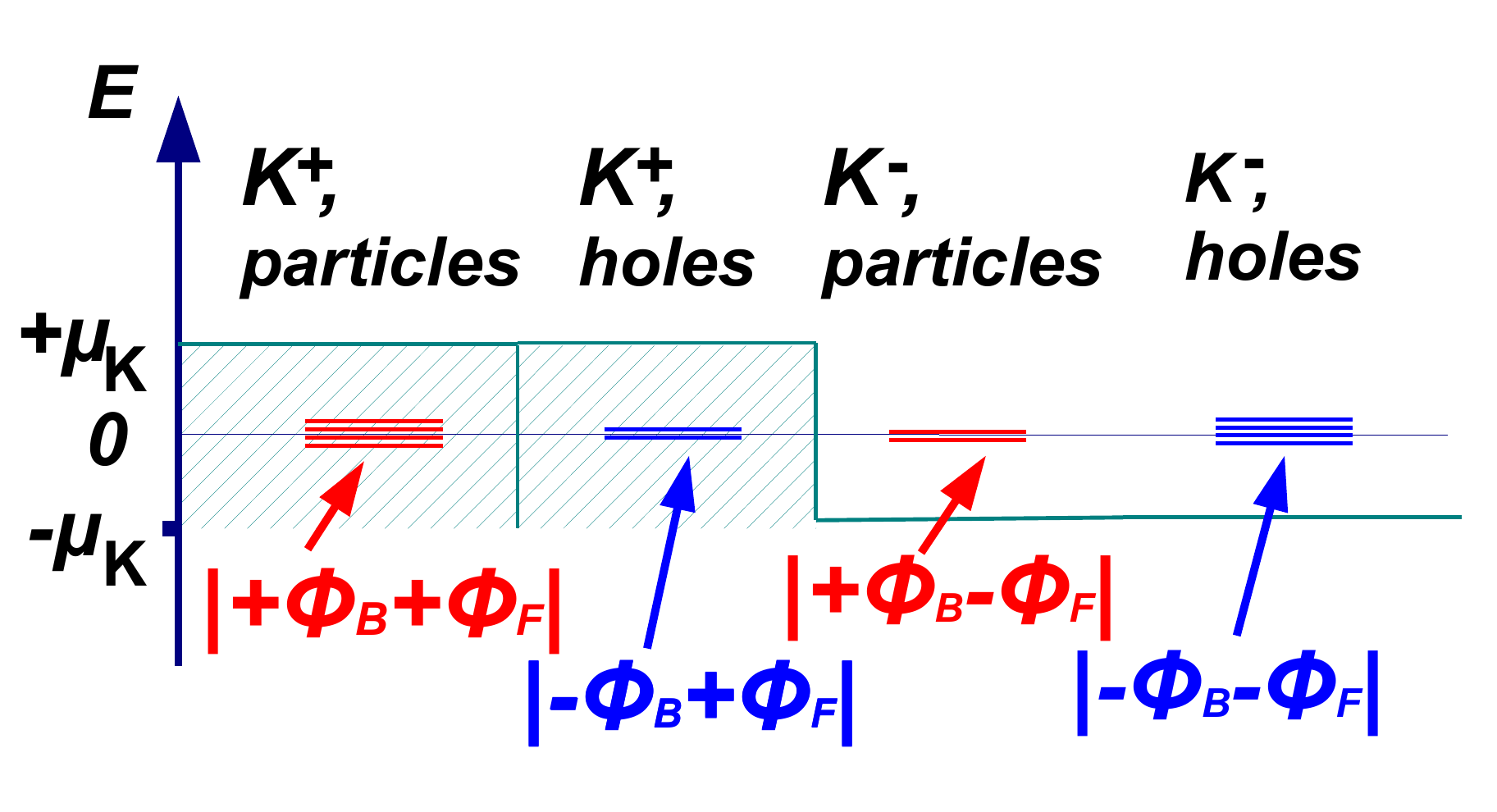}\\
  \includegraphics[width=6.4cm]{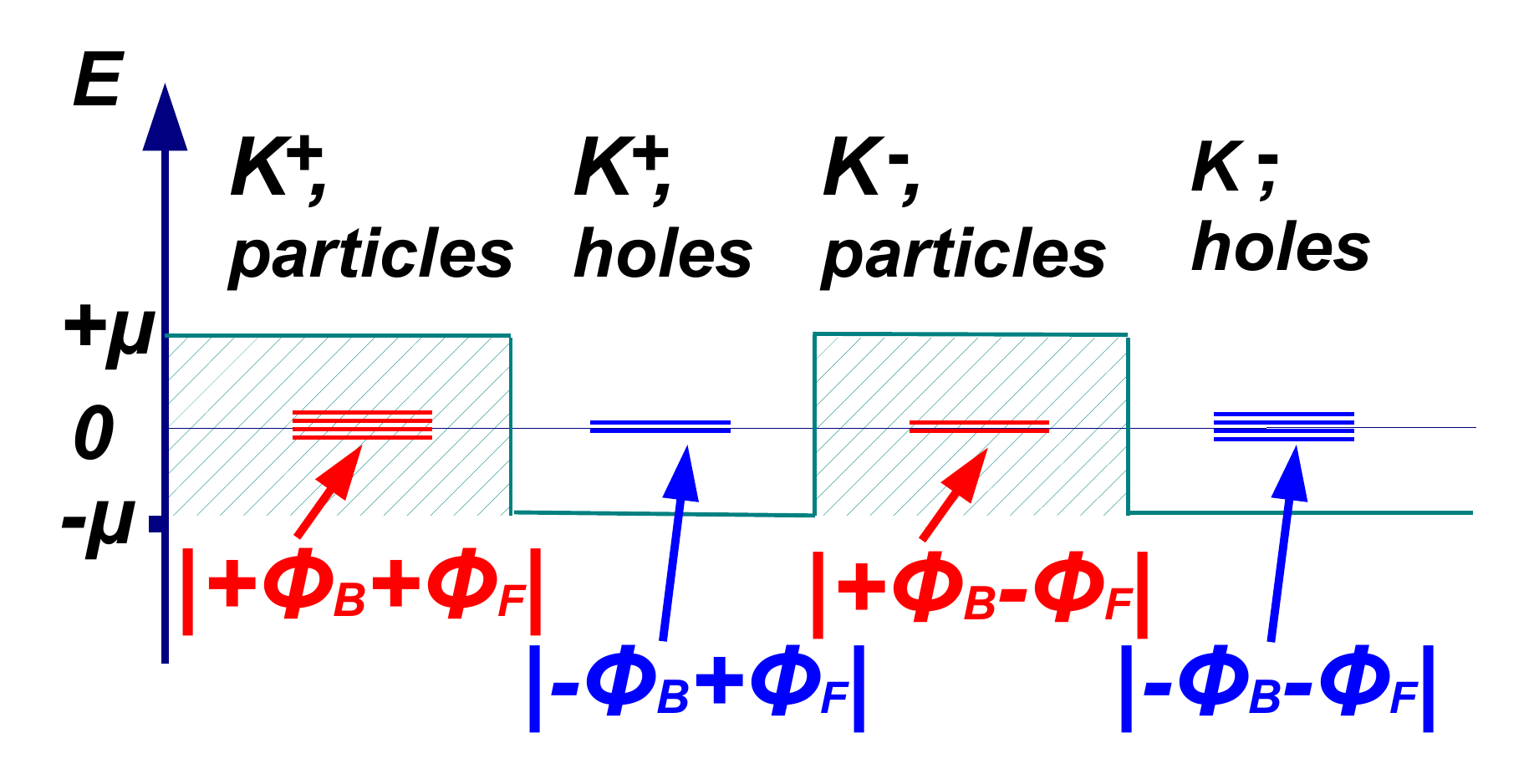}\\
  \caption{Schematic illustration of the degeneracies of the lowest Landau levels and the Fermi energies for particles and holes in different valleys in the graphene Brillouin zone at nonzero ``valley'' chemical potential (top panel) and at  nonzero chemical potential (bottom panel). As in Fig.~\ref{fig:levels_shift_general}, the number of short horizontal lines in each column denotes the degeneracy of the lowest Landau level for the corresponding quantum numbers and the long horizontal line denotes the zero energy level.}
  \label{fig:levels_shift_graphene}
\end{figure}

 Thus it is natural to associate the isospin quantum number with different valleys. Then we should also introduce the ``valley'' chemical potential $\mu_K$, so that the two valleys have different numbers of charge carriers in them. A practical method which can be used to generate valley currents (and thus imbalance in valley populations in some part of the sample) by cyclic application of strain and chemical potential has been discussed recently in \cite{Katsnelson:13:1}. Note that in contrast to the general derivation given above, the degeneracies of the lowest Landau levels in such a setting  are given by $|q B + \tau_3 F|$ and the chemical potential couples to $\tau_3$ rather than to $q \tau_3$ (see Fig.~\ref{fig:levels_shift_graphene}, top panel), but the principle by which the charge is generated is the same. Namely, now the lowest Landau levels for particles and holes in the valley $K^{+}$ are populated. Again, the level degeneracies for particles and holes in this valley are different and given by $|\Phi_B + \Phi_F|$ and $|-\Phi_B + \Phi_F|$, respectively. Thus there are more particles than holes, and the graphene sample should acquire nonzero electric charge.

 Considering the realistic experimental situation in which the strength of the magnetic field is much smaller than the strength of the pseudo-magnetic field \cite{Crommie:10:1}, the charge density per unit area can be estimated from (\ref{charge_generated}) as $Q/\lr{S B} \sim 10^{11} \, \rm{cm^{-2} T^{-1}}$. For sufficiently strong magnetic fields of order of $1 \, {\rm T}$, such charge density can be easily measured using modern experimental methods such as scanning tunnelling spectroscopy.

 Again we can invert the mechanism of charge accumulation: if instead of the valley chemical potential the ordinary chemical potential is nonzero, the lowest Landau levels will only be populated for particle states in both $K^{+}$ and $K^{-}$, as illustrated in Fig.~\ref{fig:levels_shift_graphene} at the bottom. The degeneracies of these levels are given by $|\Phi_B + \Phi_F|$ and $|\Phi_B - \Phi_F|$. Thus there are more particles in $K^{+}$ than in $K^{-}$, and the ``valley charge'' with density $|Q_{K^+} - Q_{K^-}|/\lr{S B} \sim 10^{11} \, \rm{cm^{-2} T^{-1}}$ is generated. Estimating the value of charge density in the valley pumping device of \cite{Katsnelson:13:1} as $|Q_{K^+} - Q_{K^-}|/S \sim 10^8 \, {\rm cm^{-2}}$ (the ratio of the estimated ``valley charge'' per pumping cycle to a typical area of the device), we see that for $B \sim 1 \, {\rm T}$ the magnetically induced valley asymmetry is much larger. Thus the described phenomenon can be also potentially useful for applications in (at present hypothetical) valley-based electronics \cite{Katsnelson:13:1}. We note that the possibility of breaking the symmetry between the two valleys due to combined effect of mechanical strain and magnetic field has also been discussed in \cite{Guinea:08:1}, but the possibility of charge catalysis was not considered.

\section{Hot QCD in magnetic field}
\label{sec:qcd}

 In QCD, we associate the isospin quantum number with the color of quarks, and the isospin magnetic field with the chromomagnetic field. The external magnetic field will then change the degeneracies of energy levels in different color sectors, as in Eq.~(\ref{charge_generated}). Recently, it was demonstrated that this difference of level degeneracies leads to the appearance of nonzero electric dipole moment in the background of topologically nontrivial configurations of non-Abelian gauge fields \cite{Dunne:11:1}. However, the existence of the ``color chemical potential'', which would be required to produce electric charge in this situation, is clearly prohibited by gauge invariance.

 Still, the weights of different Landau levels in the partition function can be different for different color sectors if the gauge field configuration in the Euclidean path integral representation of the thermal partition function has a nontrivial non-Abelian holonomy (Polyakov loop)
\begin{equation}
\label{polloop}
 P\lr{\vec{x}} = \mathcal{P} \expa{i \int\limits_{0}^{\beta} dx^0 A_{0}^{a}\lr{x^0, \vec{x}} \tau_a }
\end{equation}
around the compact Euclidean time direction  $x^0 \in \lrs{0, \beta}$. Here $\mathcal{P}$ is the path-ordering operator, $\beta = T^{-1}$ is the inverse temperature and $\tau_a$ are the generators of the gauge group. The time-like component of the gauge field $A_{0}^{a}\lr{x^0, \vec{x}}$ couples to quarks in a way similar to imaginary ``color'' chemical potential. This is easiest to see after diagonalizing $P$, which then does not couple the states with $\tau_3 = +1$ and $\tau_3 = -1$ anymore. However, in the Euclidean path integral $A_0$ plays the role of the Lagrange multiplier for the Gauss law and thus should be integrated over. By virtue of charge conjugation invariance the net electric charge is clearly zero after such integration. Nevertheless, as we demonstrate below, the ``virtual'' charge which appears due to non-trivial holonomy still manifests itself as a negative contribution to the \emph{fluctuations} of electric charge.

\subsection{Charge catalysis in constant chromomagnetic fields and the effect of holonomy}
\label{subsec:const_field_qcd}

 As a model assumption, let us consider the configuration of $SU\lr{2} \otimes U\lr{1}$ gauge fields in $\lr{3 + 1}$ dimensions with uniform and parallel magnetic and chromomagnetic fields with strengths $B \equiv B_{xy}$ and $F \equiv F_{xy}^3$ and a nontrivial holonomy (Polyakov loop) $P = e^{i \beta v \tau_3/2}$ which corresponds to a constant time-like component of non-Abelian gauge field:
\begin{equation}
 A_{0}^a\lr{x^0, \vec{x}} \tau_a = \frac{v}{2} \, \tau_3\,.
\end{equation}
In the language of spontaneous symmetry breaking, $v$ plays the role of a Higgs vacuum expectation value.

Such a field configuration can be thought of as an approximation of the gauge field in the so-called caloron solution, which is the topologically nontrivial saddle point of Euclidean path integral of finite-temperature gauge theories and for gauge group $SU\lr{2}$ consists of a pair of non-Abelian monopole and anti-monopole \cite{Kraan:98:2, Lee:98:1} \footnote{The approximation considered here would also work for the region between the monopole and the anti-monopole belonging to the caloron and anti-caloron, respectively, as well as for quasi-abelian monopole - anti-monopole pairs which appear in the models of QCD vacuum with Abelian dominance \cite{tHooft:81:1}.}. The chromomagnetic field between them can be roughly approximated as constant. While direct numerical studies of induced charge in the caloron background will be described in the next Subsection \ref{subsec:caloron}, here we use this simple approximation to illustrate the basic features of charge catalysis.

Assuming periodic boundary conditions in spatial directions, the fermion partition function for massless quarks in such a background can be written as
\begin{eqnarray}
\label{qcd_pf}
 \log\mathcal{Z}_f\lr{v, \mu}/V = \int \frac{dk_z}{2 \pi}
 \sum\limits_{n=0}^{+\infty} g_n \times \qquad\qquad\qquad{}
 \nonumber \\
 \left\{
  \frac{|B + F|}{2\pi} \, \log\lr{1 + e^{-\beta E^{+}_{n,\, k_z} + i \beta v/2 + \beta \mu}}
 \right.
  + \nonumber \\ +
  \frac{|-B - F|}{2\pi} \, \log\lr{1 + e^{-\beta E^{+}_{n,\, k_z} - i \beta v/2 - \beta \mu}}
  + \nonumber \\ +
  \frac{|B - F|}{2\pi} \, \log\lr{1 + e^{-\beta E^{-}_{n,\, k_z} - i \beta v/2 + \beta \mu}}
  + \nonumber \\ +
  \left.
  \frac{|-B + F|}{2\pi} \, \log\lr{1 + e^{-\beta E^{-}_{n,\, k_z} + i \beta v/2 - \beta \mu}}
 \right\} ,
\end{eqnarray}
where $k_z$ is the momentum along the magnetic fields, $V$ is the total spatial volume and
\begin{equation}
 E^{\pm}_{n,\, k_z} = \sqrt{2 |B \pm F| n + k_z^2}
\end{equation}
are the Landau levels for particle/anti-particle states with $\tau_3 = \pm 1$. We have also introduced an infinitesimally small chemical potential $\mu$ as a source for electric charge. In the end it will be set to zero. The the third and the fourth lines in (\ref{qcd_pf}) correspond to the particle and anti-particle states in the $\tau_3 = +1$ sector, while the fifth and the sixth line are the same for the $\tau_3 = -1$ sector. To simplify the discussion further, let us assume that the temperature is sufficiently small ($T^2 \ll \mymin{|F-B|, |F+B|}$), so that the lowest Landau levels with $E^{\pm}_{0, k_z} = |k_z|$ dominate in (\ref{qcd_pf}). Calculating now the derivative of (\ref{qcd_pf}) with respect to $\mu$ at $\mu = 0$ and evaluating the resulting integrals over $k_z$ analytically, we obtain for the expectation value of the charge density over the fermion states:
\begin{align}
\label{charge_density}
 \vev{q}_f = &\, \frac{T}{V}\,\mathcal{Z}_f^{-1}\lr{v, \mu}\frac{\partial}{\partial \mu} \mathcal{Z}_f\lr{v, \mu}|_{\mu \rightarrow 0}\\
  = &\,\frac{i v}{\lr{2 \pi}^2} \, \lr{|B + F| - |B - F|}, \qquad \beta v \in \lrs{-2 \pi, 2 \pi}\nonumber
\end{align}
We also note that the Polyakov loop (\ref{polloop}), the partition function (\ref{qcd_pf}) and the fermionic expectation value of a charge (\ref{charge_density}) are periodic functions of $v$ with the period $4 \pi T$, thus $v$ is treated as a compact variable with $v \in \lrs{-2 \pi T ; 2 \pi T}$.

 This result is very similar in form to (\ref{charge_generated}), except for the factor of $i$ in front. The origin of this factor is clear from (\ref{qcd_pf}): the holonomy plays the role of imaginary chemical potential for the $\tau_3$ quantum number and thus induces imaginary electric charge. However, in contrast to (\ref{charge_generated}), the quantity (\ref{charge_density}) does not correspond to the expectation value of a Hermitian operator in the ground state of some Hamiltonian. The correspondence with the canonical formalism is only recovered after integration over the holonomy, which obviously results in zero expectation value of the physical charge operator. Consider, however, the fermionic expectation value of the squared charge:
\begin{align}
\label{charge_squared}
 \vev{Q^2}_f =  &\,T^2 \, \mathcal{Z}_f^{-1}\lr{v, \mu} \,
 \frac{\partial^2}{\partial \mu^2} \mathcal{Z}_f\lr{v, \mu}|_{\mu \rightarrow 0}
 \\
 = &\,
 \frac{V T \, \mymax{F, B}}{\pi^2} - \frac{4 v^2 \, V^2 \, \mymin{B^2, F^2}}{\lr{2 \pi}^4} \nonumber .
\end{align}

 The first summand is the contribution of connected fermion diagram and is independent of the holonomy. The second term corresponds to disconnected fermion diagrams and is exactly the square of the ``virtual'' charge given by (\ref{charge_density}). We see thus that the effect of holonomy is to reduce the charge fluctuations. We note that a suppression effect of nontrivial holonomy on charge density at finite chemical potential (without magnetic fields) has also been found in \cite{Langfeld:10:1}.

 Finally, we have to integrate the above expression over all values of the holonomy with the weight $\mathcal{Z}_f\lr{v,\mu = 0}$. Assuming again that the lowest Landau level gives the dominant contribution to the partition function and performing the integration over $k_z$ in (\ref{qcd_pf}), we obtain
\begin{eqnarray}
\label{pf_LLL}
\mathcal{Z}_f\lr{v,\mu = 0} = \const \times \expa{-\frac{V \mymax{|F|, |B|}}{8 \pi^2 T} \, v^2} .
\end{eqnarray}
Approximating the compact integral over $v \in \lrs{-2 \pi T, 2 \pi T}$ by a Gaussian integral over the whole real axis (which makes sense for sufficiently large volumes), we obtain the following result for the expectation value of the squared charge
\begin{eqnarray}
\label{ChargeFluctUniformFinal}
 \vev{Q^2} = \frac{V T \, \mymax{|F|, |B|}}{\pi^2} \, \lr{1 - \frac{\mymin{F^2, B^2}}{\mymax{F^2, B^2}}} .
\end{eqnarray}
Thus charge fluctuations become suppressed as the magnetic field grows until the strength of chromomagnetic and magnetic fields become equal. At that point the fluctuations become zero, and then start growing again.

 Note also that for the integration over the holonomy in (\ref{ChargeFluctUniformFinal}) we did not use the $SU\lr{2}$ Haar measure $d v \sin^2\lr{\frac{v}{2 T}}$, but rather the $U\lr{1}$ Haar measure on the interval $\lrs{-2 \pi T; 2 \pi T}$. The inclusion of the full Haar measure would be justified only if we considered the full path integral over the gauge fields. Indeed, one can easily check that the expectation value $\vev{Q^2}$ calculated with the $SU\lr{2}$ Haar measure becomes negative if $|F|/|B|$ is close to unity. Since $\vev{Q^2}$ is the expectation value of the square of the Hermitian operator and should be manifestly positive, we conclude that the inclusion of the Haar measure alone leads to an inconsistent result. On the other hand, one can try to take into account the gaussian fluctuations of the gauge fields around the background field $F$. In this case the contribution of the fluctuations of the longitudinal components of the gauge field to the Polyakov loop effective action exactly cancels the Haar measure contribution (see, for example, equation (22) in \cite{Weiss:81:1}). Thus the positivity of $\vev{Q^2}$ would be recovered if we self-consistently improved our calculation by including both the $SU\lr{2}$ Haar measure and the fluctuations of the background field (in the gaussian approximation). In view of this observation, it seems more natural to use the flat $U\lr{1}$ measure in case one considers only a constant background field configuration without taking into account any fluctuations on top of it. In any case, our main qualitative conclusion - the decrease of charge fluctuations with increasing magnetic field - will not change if we modify the integration weight for $v$.

\subsection{Electric charge and current on caloron configurations}
\label{subsec:caloron}

\begin{figure}
  \includegraphics[width=8.5cm]{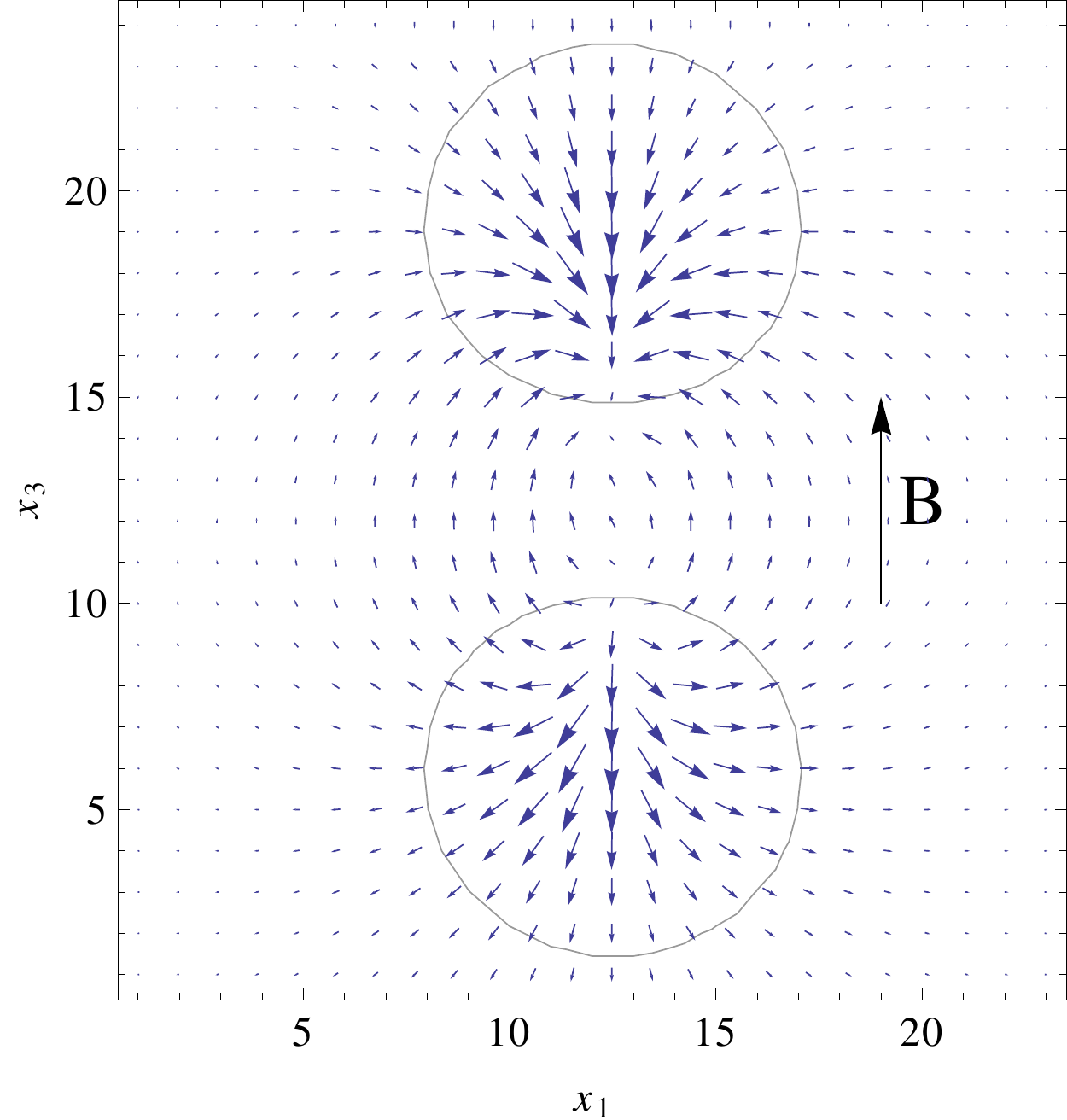}\\
  \caption{Characteristic gluonic quantities of an $SU(2)$ caloron with maximally nontrivial holonomy on a $24^3\times 6$ lattice in two spatial dimensions with monopoles separated along $x_3$. The third color component of the chromomagnetic field is depicted by short blue arrows. We also show the direction of the Abelian magnetic field (for the case when it is parallel to the caloron symmetry axis) with a long black arrow. The Polyakov loop is characterized by its contour lines $1/2\cdot \tr P=\pm0.4$. At the monopole/anti-monopole the Polyakov loop becomes trivial $1/2\cdot \tr P\to \pm1$, while outside of the monopoles the Polyakov loop becomes diagonal (proportional to $\tau_3$) and traceless, $1/2\cdot \tr P\to 0,\, v\to \pi T$.}
  \label{fig:mag_caloron}
\end{figure}

 As a more realistic model of quantum fluctuations in non-Abelian gauge theory one can consider a gas of topologically nontrivial gauge field configurations, such as an instanton gas \cite{Shuryak:98:1}. However, at finite temperature instantons are no longer the solutions of classical Euclidean Yang-Mills equations. A generalization of the instanton solution to Euclidean space with compact time direction is the caloron solution, either with a trivial \cite{Harrington:78:1} or nontrivial \cite{Kraan:98:1, Kraan:98:2, Kraan:98:3, Lee:98:1, Lee:98:2} holonomy at spatial infinity. The former is stable at sufficiently high temperatures, while the latter is believed to dominate at lower temperatures \cite{Ilgenfritz:07:1}. Much like instantons, calorons are strongly localized solutions. For $SU(2)$ gauge group, they consist of a pair of Bogomol'ny-Prasad-Sommerfeld monopole and anti-monopole, separated by some distance. In the region of space between the monopole and the anti-monopole the chromomagnetic and chromoelectric fields can be roughly approximated as constant and parallel to the caloron axis, see Fig.~\ref{fig:mag_caloron}. Hence, if the magnetic field is also parallel to the caloron axis, such solution can realize the case of parallel magnetic and chromomagnetic fields described above. We can thus expect that the magnetic field superimposed on the caloron configuration leads to the appearance of nonzero (imaginary) electric charge.

 Explicit expressions for the Dirac propagator in the combined background of a caloron plus magnetic field are not known, therefore we have to find it numerically. To this end, we put the $SU\lr{2}$ caloron field configurations on the lattice. The explicit form of the caloron field configuration which we use is given by the expression (57) in \cite{Kraan:98:2}. We use calorons of maximally nontrivial holonomy meaning that the asymptotic Polyakov loop becomes traceless (i.e.\ $v= \pi T$ asymptotically), for which the monopole and anti-monopole have equal topological charges $1/2$. Lattice discretization is done by assigning each link variable with the path-ordered exponent of the continuum gauge field $A_{\mu}^a\lr{x} \tau_a$ along the link. More details on the lattice discretization of the caloron solutions can be found in \cite{Ilgenfritz:07:1, Ilgenfritz:07:2, Bruckmann:10:1, Rodl:13:1}. To keep discretization errors under control, we have also performed calculations for field configurations obtained from the initial caloron configuration after several cooling steps. We have found that cooling makes no significant effect on the generated charge, thus discretization artifacts should be quite small.

 Since we are studying the phenomena closely related to quark chirality (in particular, zero quark mass is important to ensure the existence of the lowest Landau level of zero energy), we use the chirally invariant Neuberger's overlap Dirac operator \cite{Neuberger:98:1} to calculate the vacuum expectation value of the current. Since the caloron configuration has a unit topological charge, the Dirac operator has exactly one zero mode, which makes it formally non-invertible in the massless limit. To regulate this trivial divergence, we introduce a small quark mass $m_q = 0.01$ in lattice units. Then the Dirac operator is inverted using the SHUMR method \cite{Borici:06:1, Borici:06:2}. In order to check that our results are not affected by finite-volume and boundary effects, we consider caloron configurations of the same size on $16^3 \times 4$, $20^3 \times 4$ and $24^3 \times 4$ lattices. The dependence on the lattice spacing is checked by rescaling the caloron configuration that has been put on the $16^3 \times 4$ lattice by a factor $3/2$ and putting it on the $24^3 \times 6$ lattice. Thus we assume that the lattice spacing is $a = 1$ for the $16^3 \times 4$, $20^3 \times 4$ and $24^3 \times 4$ lattices, and $a = 2/3$ for the $24^3 \times 6$ lattice. Since the current and charge densities are operators of canonical dimension $3$, their expectation values are also multiplied by $\lr{3/2}^3$. Similarly, the magnetic field strength is rescaled by a factor $\lr{3/2}^2$. Note that the magnetic flux is a dimensionless quantity which is not changed when we simultaneously increase the lattice size by some factor and decrease the lattice spacing by the same factor. Of course, there is no intrinsic physical dimensionful scale in our problem, since we deal with a fixed gauge field configuration. The physical scale can be restored, for example, from the characteristic size of calorons in the phenomenological model of the caloron gas \cite{Ilgenfritz:07:1}. This will be done at the end of the paper, where we will discuss possible experimental consequences of our results.

\begin{figure*}[htpb]
  \includegraphics[width=6cm, angle=-90]{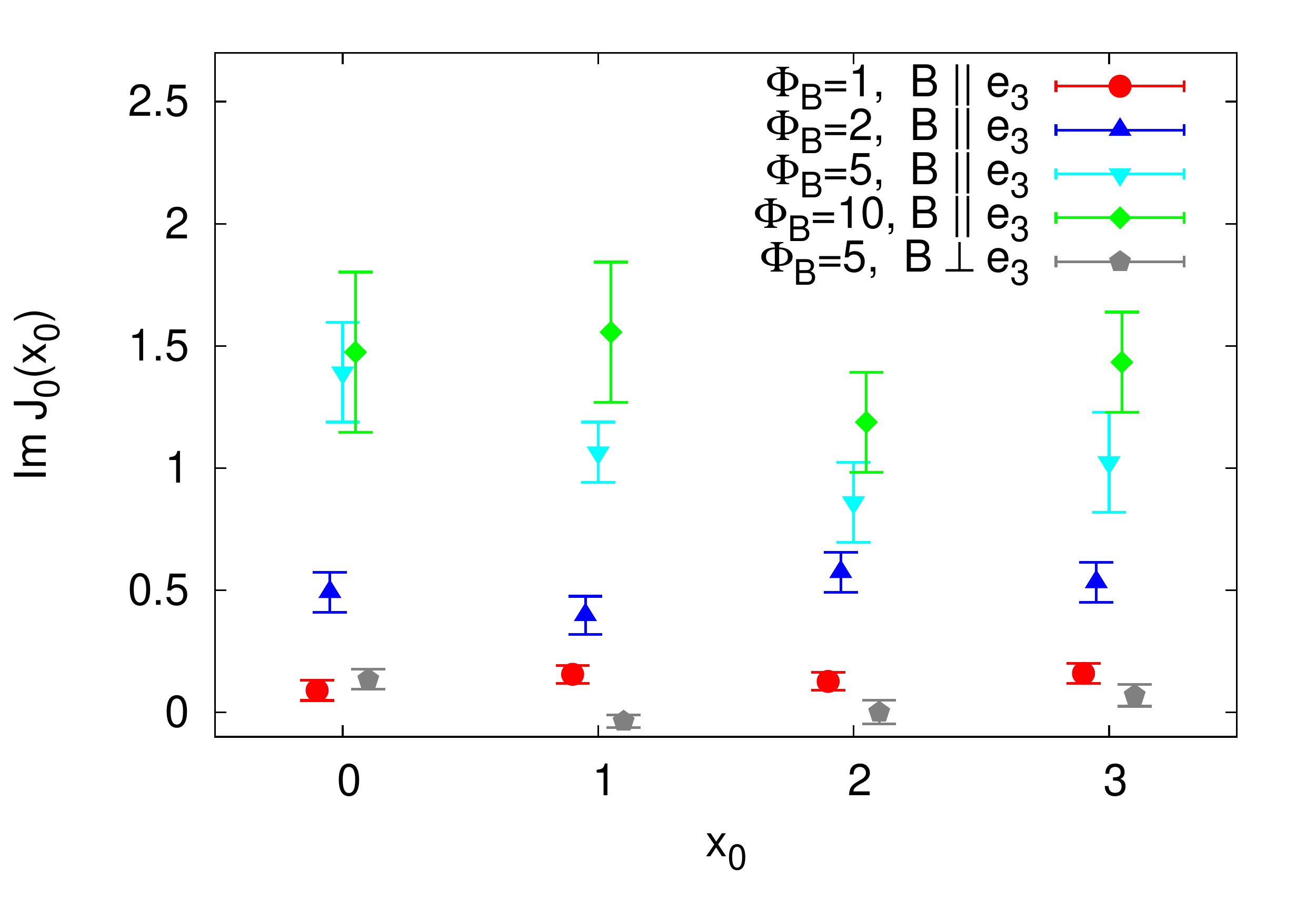}
  \includegraphics[width=6cm, angle=-90]{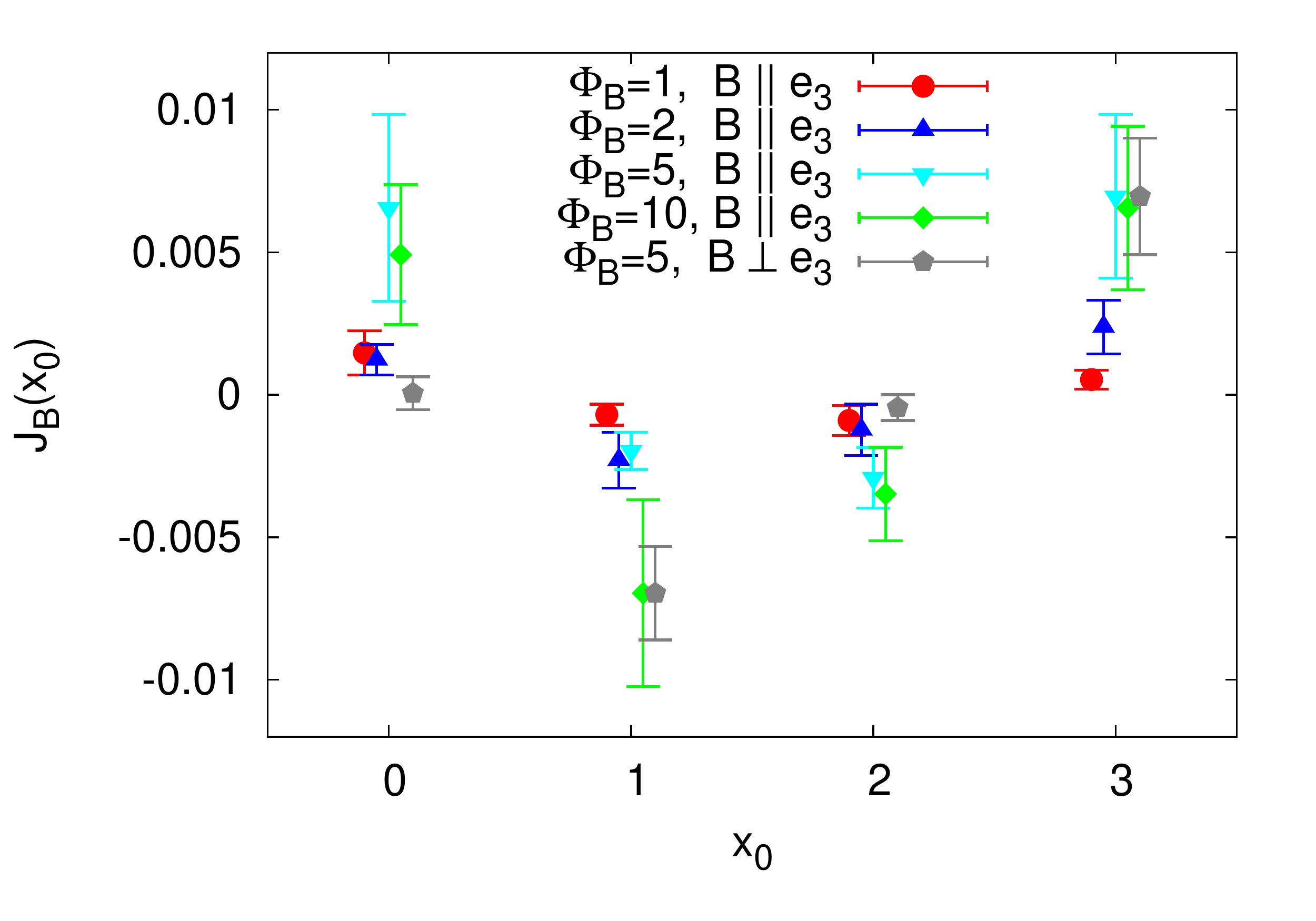}\\
  \caption{Imaginary part of the net electric charge $J_0\lr{x_0}$ (left) and the net electric current in the direction of the magnetic field $J_B\lr{x_0}$ (right) as a function of the discrete lattice time $x_0$ on the $16^3 \times 4$ lattice. The data points are slightly displaced along the horizontal axis in order to make the plots more illustrative.}
  \label{fig:J0_Jz}
\end{figure*}

 To start with, we consider the imaginary part of the net electric charge $J_0\lr{x_0}$ and the net electric current $J_B\lr{x_0}$ in the direction of the magnetic field $\vec{B}$ as a function of Euclidean lattice time $x_0$. Spatial components of the current are manifestly real, since in Euclidean space they are still given by functional derivatives of the free energy over the gauge field $A_{\mu}\lr{x}$:
\begin{align}
\label{CurrentVEV}
 \vev{ j_i\lr{x} }_f &\,\sim \frac{\delta}{\delta A_{i}\lr{x}} \log\det{\mathcal{D}\lrs{A_{\mu}\lr{x}}  }
 \nonumber\\ &\,\sim
 -i \tr\lr{\mathcal{D}^{-1}\lrs{A_{\mu}\lr{x}} \gamma_{i}}, \quad i = 1, 2, 3,
\end{align}
where by $\vev{\ldots}_f$ we again denote the expectation value with respect to the fermion fields only. Since the Dirac operator $\mathcal{D}\lrs{A_{\mu}\lr{x}} = \gamma^{\mu} \lr{\partial_{\mu} - i A_{\mu}\lr{x}}$ is anti-Hermitian and Euclidean gamma-matrices are Hermitian, the resulting expression is real. On the other hand, the expectation value of the charge is the derivative of the partition function with respect to the chemical potential:
\begin{align}
\label{ChargeVEV}
 \vev{ Q }_f &\, \sim  \frac{\partial}{\partial \mu} \log\det{\mathcal{D}\lrs{A_{\mu}\lr{x}} + \mu \gamma_0 }|_{\mu \rightarrow 0}
 \nonumber \\
&\,\sim
  \tr\lr{\mathcal{D}^{-1}\lrs{A_{\mu}\lr{x}} \gamma_{0}} ,
\end{align}
for which the real part is identically zero at $\mu = 0$. As discussed above in Subsection \ref{subsec:const_field_qcd}, this imaginary charge should vanish after functional integration over the time-like component of the gauge field $A_0\lr{x}$.

 The net charge and current are obtained by summing the local charge and current densities over fixed time slices in Euclidean space:
\begin{align}
\label{TotalCharge}
 J_0\lr{x_0} = &\, \int \!d^3 x \,\vev{j_0\lr{x_0, \vec{x}}} \,,\\
 J_B\lr{x_0} = &\, \int \! d^3 x \, \vev{\vec{j}\lr{x_0, \vec{x}}} \cdot \vec{B}/|B| \,.
\end{align}
where 
\begin{eqnarray}
\label{charge_dens_def}
 \vev{j_{\mu}\lr{x_0, \vec{x}}}_f = \tr\lr{\mathcal{D}^{-1}\lr{x_0, \vec{x}; x_0, \vec{x}} \gamma_{\mu} }  .
\end{eqnarray}
We introduce this volume integration as an estimate of the macroscopic currents, which are studied in experiments at scales which are much larger than the typical correlation length of the QCD vacuum \cite{Kharzeev:08:1}.

\begin{figure}[htpb]
  \includegraphics[width=6cm, angle=-90]{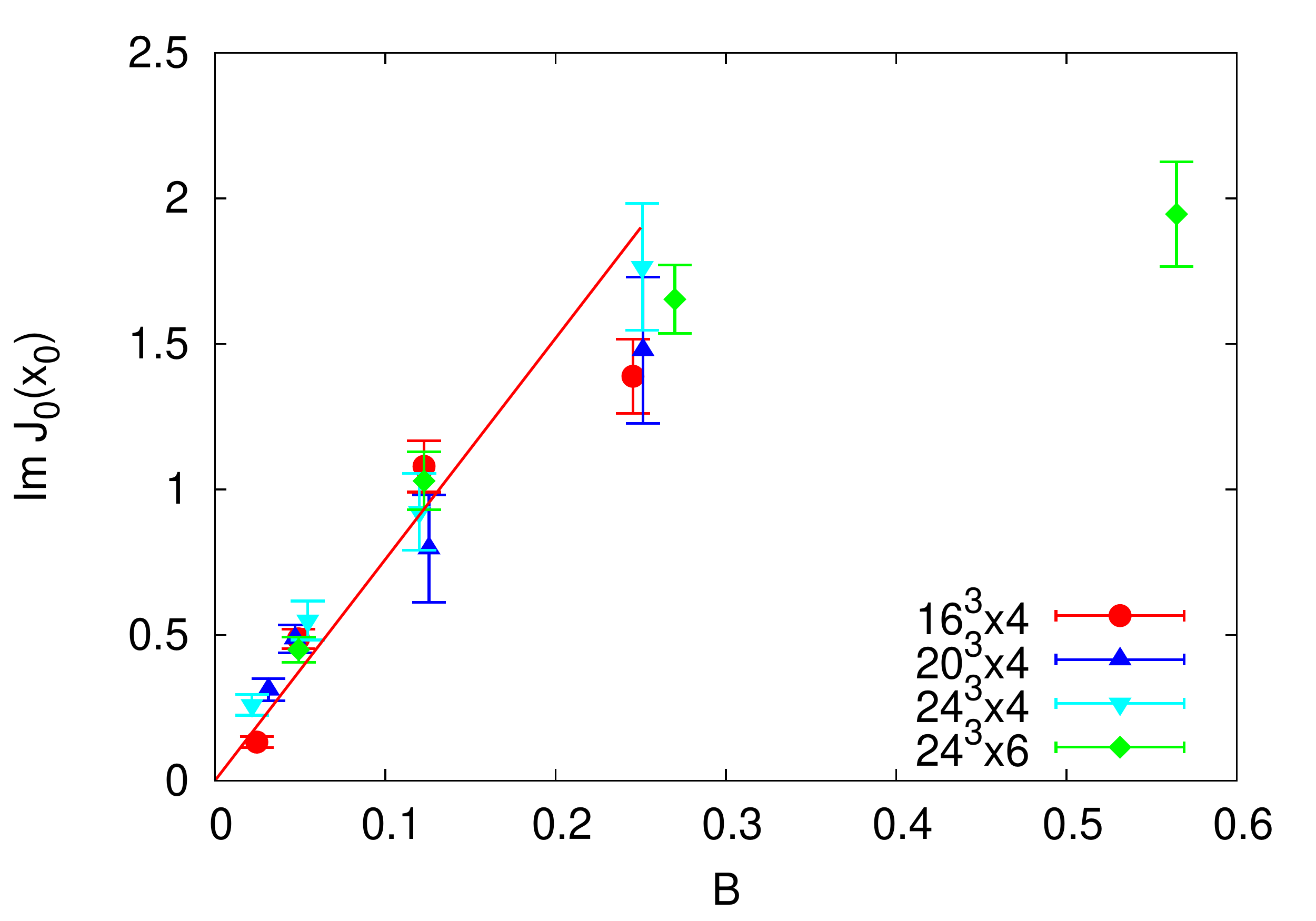}\\
  \caption{Imaginary part of the net electric charge $J_0\lr{x_0}$, averaged over all $x_0$, as a function of the magnetic field strength on different lattices.}
  \label{fig:caloron_total_charge}
\end{figure}

 In Fig.~\ref{fig:J0_Jz} we plot the net overall current $J_B\lr{x_0}$ as well as the imaginary part of the charge density $\Im J_0\lr{x_0}$ for the $16^3 \times 4$ lattice, for magnetic fields both parallel and perpendicular to the caloron symmetry axis (which is parallel to the $\vec{e}_3$ basis vector) and for different field strengths. Despite the fact that we use a fixed and sufficiently smooth gauge field configuration, the inversion of the overlap Dirac operator still takes a lot of computer time, and the calculation of the Dirac propagator $D^{-1}\lr{x, x}$ in (\ref{ChargeVEV}) and (\ref{CurrentVEV}) for all lattice sites would be prohibitively expensive. To overcome this difficulty, we perform the integration over the spatial coordinates $\vec{x}$ in the definitions of $J_0\lr{x_0}$ and $J_B\lr{x_0}$ by a Monte-Carlo method, spreading the points $\vec{x}$ uniformly over time slices and summing up the local current densities in these points. We use from $50$ to $100$ random points, depending on the lattice size. Error bars on Figs. \ref{fig:J0_Jz} and \ref{fig:caloron_total_charge} illustrate the estimated uncertainties of the Monte-Carlo integration.

 For the parallel case, one can clearly see that the imaginary part of the net electric charge grows with the field strength and practically does not depend on $x_0$. For the perpendicular case, the charge is close to zero for all $x_0$. The net current in the direction of the magnetic field $J_B\lr{x_0}$ is smaller than the imaginary part of the charge density by more than two orders of magnitude. The electric current is nonzero in both the parallel and the perpendicular cases, but shows an oscillating behavior, so that the sum over all $x_0$ is zero within the errors of Monte-Carlo integration.
Note that the caloron we use has almost static topological charge and action density, whereas the gauge fields in its constituent monopoles' cores are gauge-rotated in time with respect to each other \cite{Kraan:98:2}; this twist could cause the change of sign of the current $J_B\lr{x_0}$. Such behavior of $J_B\lr{x_0}$ suggests that no net electric charge is transported in the direction of the magnetic field in the tunneling event and thus this caloron cannot be considered as a model field configuration for the Chiral Magnetic Effect. On the other hand, the induced imaginary electric charge should contribute to the fluctuations of the net macroscopic charge, as discussed in Subsection \ref{subsec:const_field_qcd} above. In the following, we consider only the more interesting case when the magnetic field is parallel to the caloron axis (that is, parallel to $\vec{e}_3$).

\begin{figure*}[htpb]
  \includegraphics[width=6cm, angle=-90]{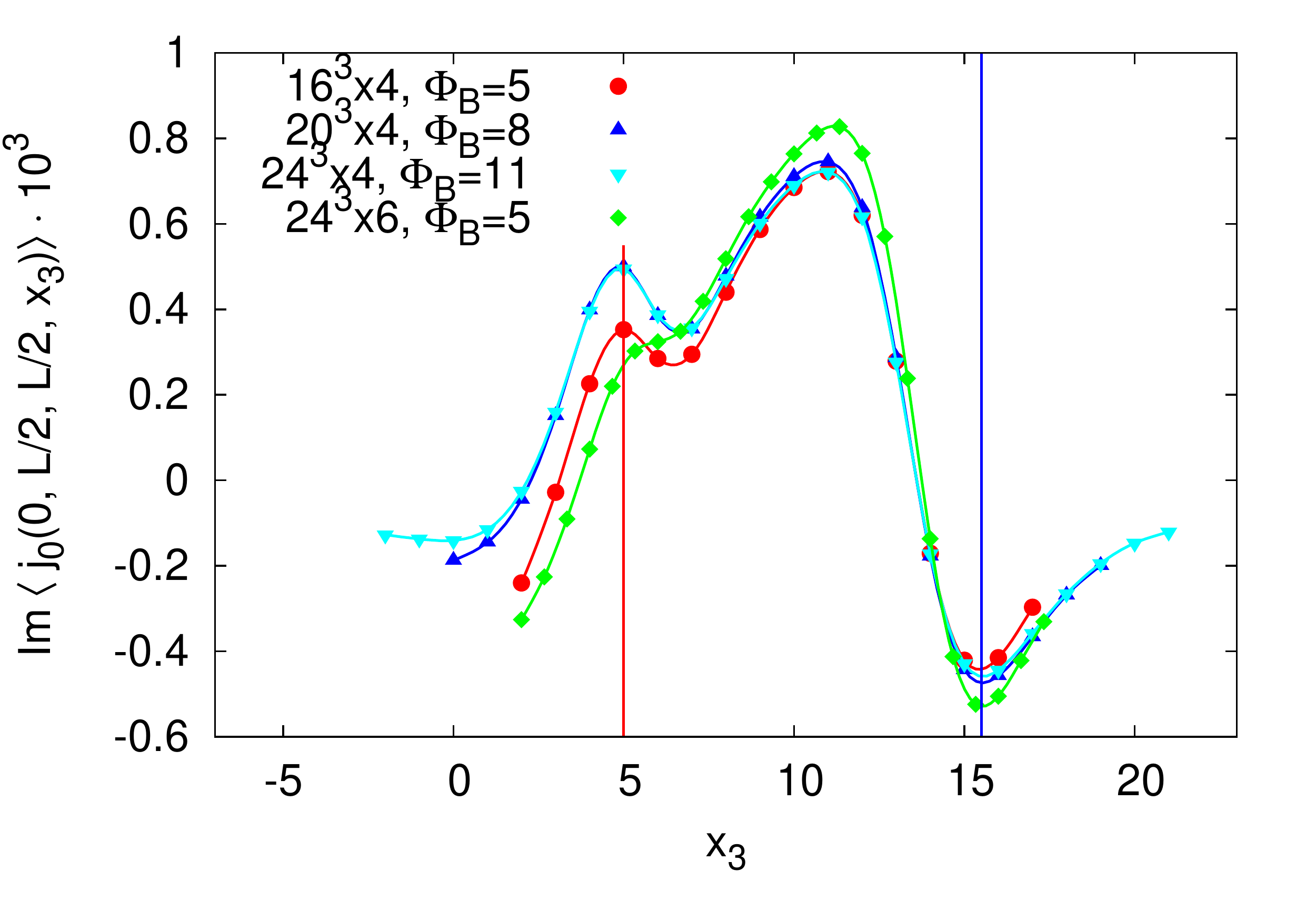}
  \includegraphics[width=6cm, angle=-90]{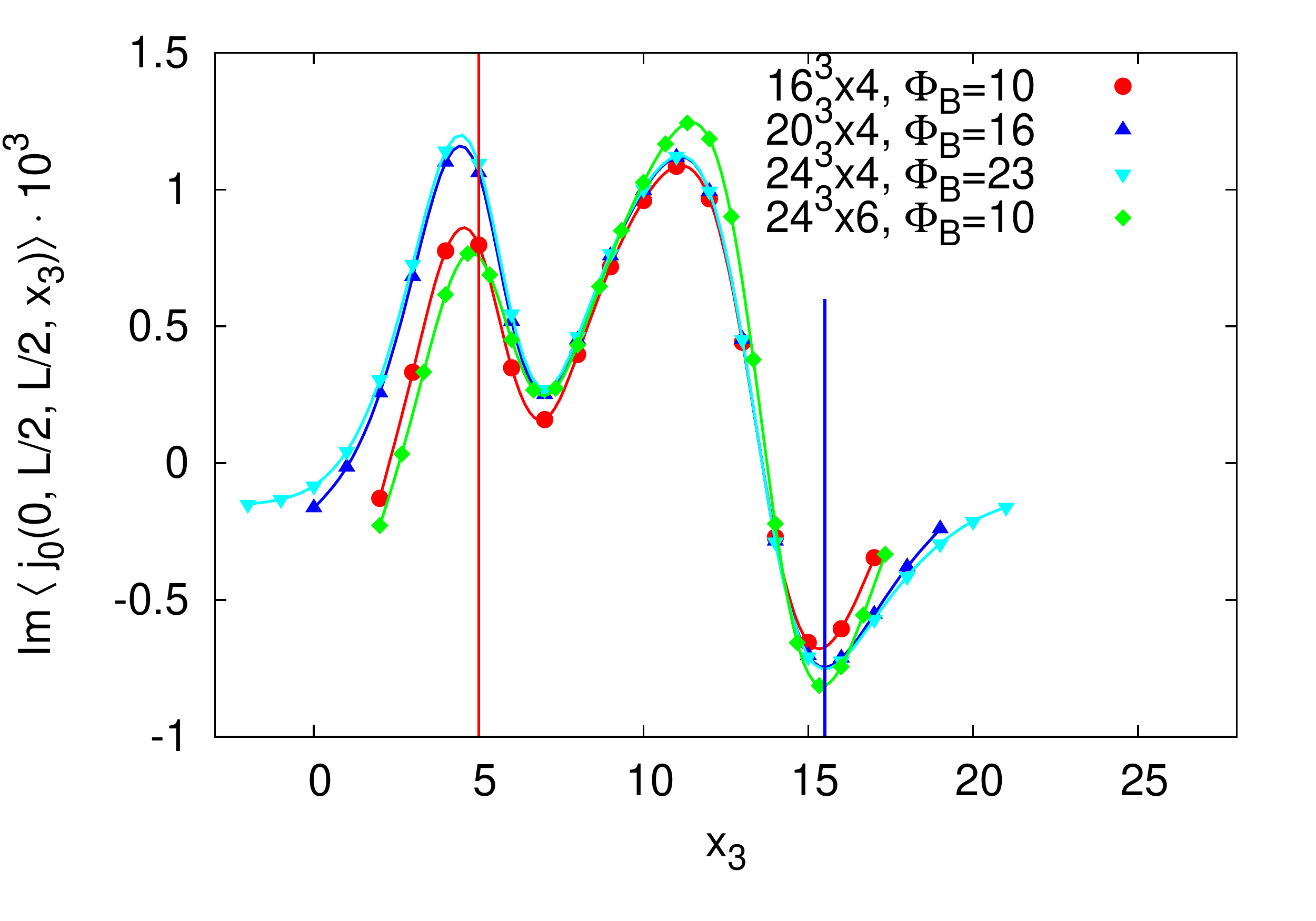}\\
  \caption{Profiles of the charge density along the symmetry axis of the caloron on lattices of different sizes for fixed values of the magnetic field strength $B$. Red and blue vertical lines mark the positions of the monopole and the anti-monopole.}
  \label{fig:charge_densities_z_vs_size}
\end{figure*}

 In Fig.~\ref{fig:caloron_total_charge} we plot the imaginary part of the net electric charge $\Im J_0\lr{x_0}$, averaged over all $x_0$, on different lattices as a function of the magnetic field strength. The value of the magnetic field strength $B$ on different lattices is kept constant (the total magnetic flux $\Phi_B$ is changed correspondingly). At moderately small field strengths the charge rises approximately linearly, and then seems to saturate at strong magnetic fields. Such behaviour is in complete agreement with the considerations of the previous Subsection \ref{subsec:const_field_qcd}. The solid line on Fig.~\ref{fig:caloron_total_charge} is a linear fit through data points with $B < 0.2$, which yields $J_0 = \lr{8 \pm 2} B$. The results obtained on different lattices agree within the uncertainties of the Monte-Carlo integration, which suggests that the induced charge is not a finite-volume nor a discretization artifact. Some specific lattice artifacts for the density of induced charge will be discussed below.

 Next we investigate the longitudinal profile of the imaginary part of the charge density $\Im \vev{j_0\lr{x}}$ along the caloron symmetry axis with $x_1 = L/2$, $x_2 = L/2$, and the transverse profile in the direction perpendicular to it. We consider the fixed time slice $x_0=0$, where the net charge is maximal. In Fig.~\ref{fig:charge_densities_z_vs_size} we show the longitudinal profiles of the charge density on lattices of different size and spacing. For the left and the right plots of Fig.~\ref{fig:charge_densities_z_vs_size}, the magnetic field strengths differ by a factor of two. The coordinate $x_3$ on the plots is shifted by $\pm 2$ or rescaled by $2/3$ in order to match the positions of the monopole and the anti-monopole on different lattices. These positions are marked by red (monopole) and blue (anti-monopole) vertical lines on Fig.~\ref{fig:charge_densities_z_vs_size}. In what follows, we denote the corresponding $x_3$ coordinates as $x_{3}^{\lr{m}}$ and $x_{3}^{\lr{a}}$, respectively. For the $16^3 \times 4$ and $24^3 \times 6$ lattices $x_{3}^{\lr{m}} = L/4$ and $x_{3}^{\lr{a}} = 3 L/4$. For the $20^3 \times 4$ and $24^3 \times 4$ lattices we simply add more space around the caloron, so that the distance between the monopole and the anti-monopole $x_{3}^{\lr{a}} - x_{3}^{\lr{m}} = L/2$ and the position of the midpoint between them $\lr{x_{3}^{\lr{a}} + x_{3}^{\lr{m}}}/2 = L/2$ are fixed.

 One can see that the charge density has a characteristic three-peak structure. Namely, there is an excess of positive imaginary charge peaked around the monopole position, and an excess of negative imaginary charge near the anti-monopole. Furthermore, the charge density has a pronounced positive peak right in the middle between the monopole and the anti-monopole, which is higher than the peaks near the monopole and the anti-monopole. The height of the middle peak becomes almost two times larger as the magnetic field strength doubles. The side peaks become also higher and more pronounced. Comparing the charge profiles for all four different lattices, we conclude that boundary effects and finite-spacing artifacts are rather small and do not exceed $20\%$.

\begin{figure*}[htpb]
  \includegraphics[width=4cm, angle=-90]{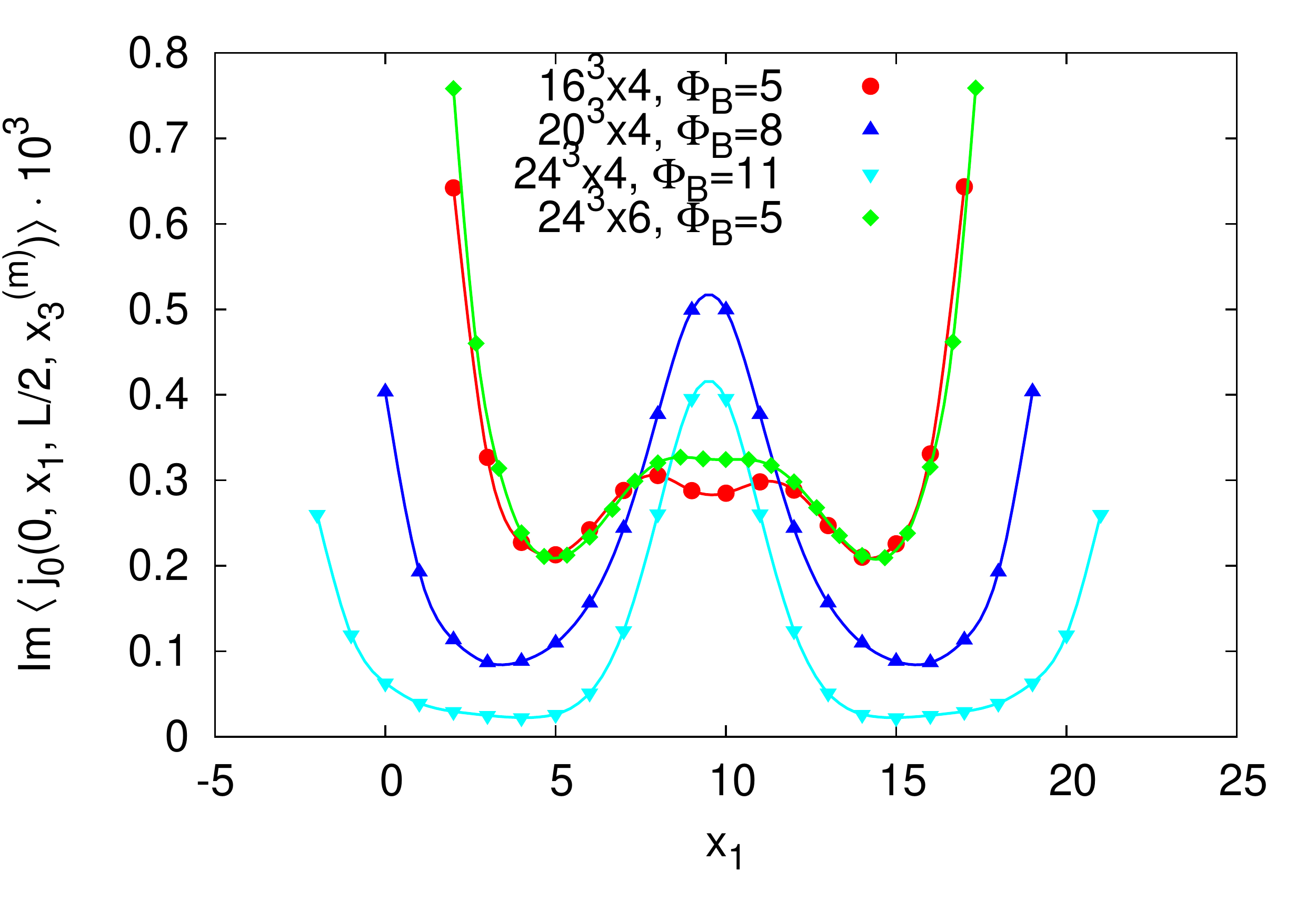}
  \includegraphics[width=4cm, angle=-90]{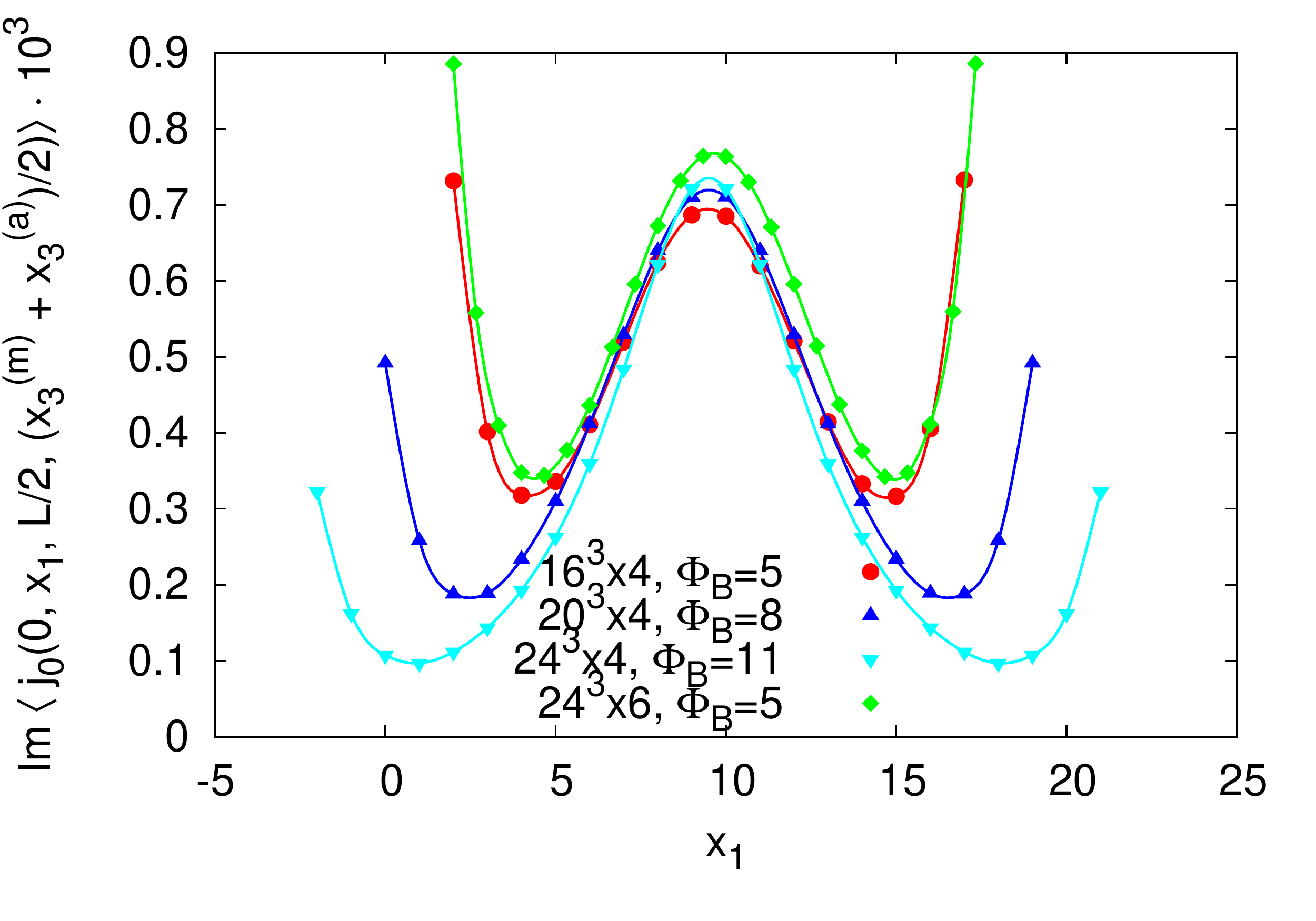}
  \includegraphics[width=4cm, angle=-90]{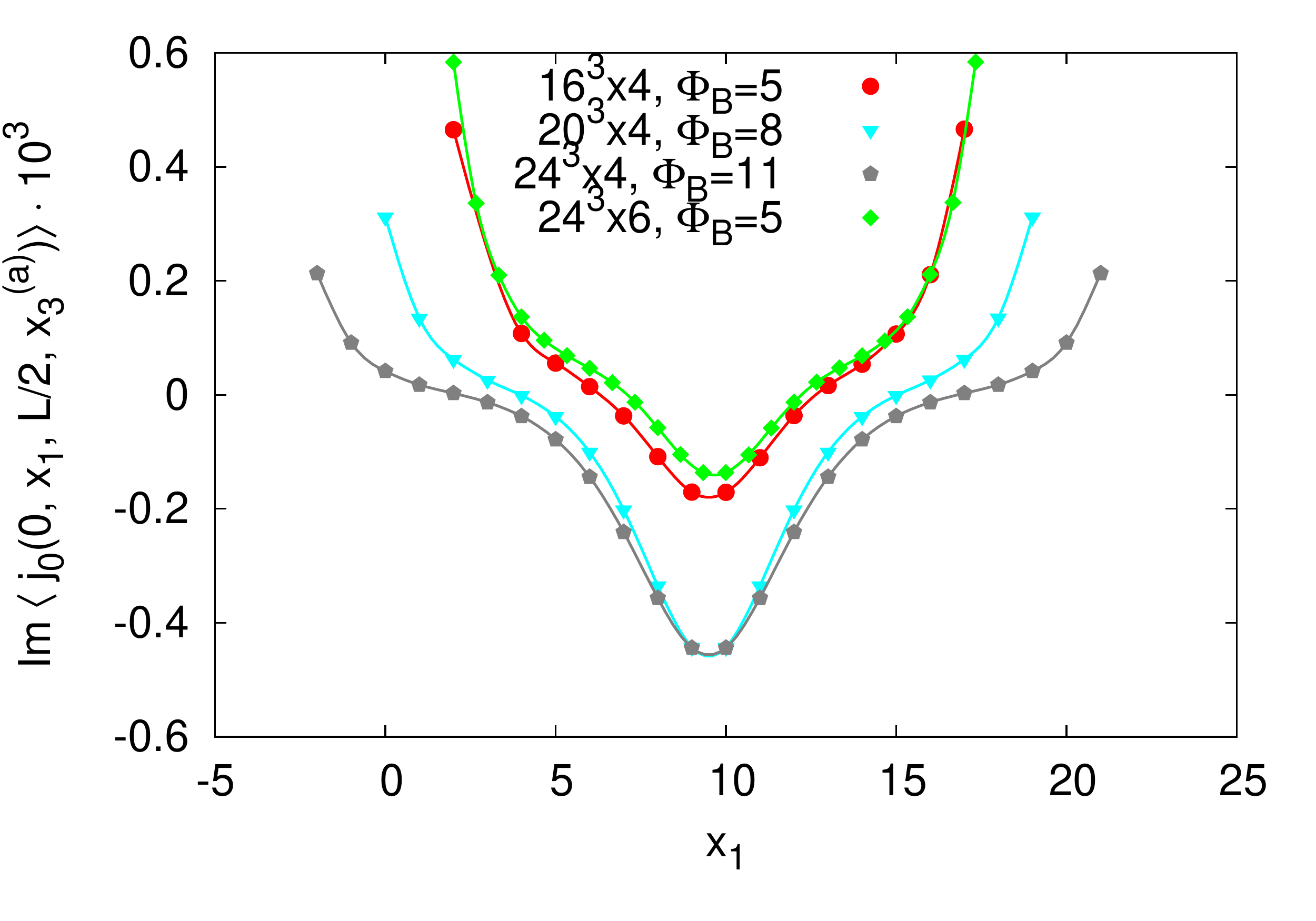}\\
  \caption{Profiles of the imaginary part of the charge density $\Im \vev{j_0\lr{x}}$ in the direction perpendicular to the symmetry axis of the caloron for different lattice parameters and at different $x_3$: at the monopole position (left), at the midpoint between the monopole and at the anti-monopole (middle) and at the anti-monopole position (right). The magnetic field strength corresponds to $\Phi_B = 5$ units of magnetic flux through the $16^3 \times 4$ lattice.}
  \label{fig:transverse_profiles}
\end{figure*}

 In order to investigate the transverse profiles of the charge density, we set $x_3 = x_3^{\lr{m}}$, $x_3 = \lr{x_3^{\lr{m}} + x_3^{\lr{a}}}/2$ and $x_3 = x_3^{\lr{a}}$ and change the transverse coordinate $x_1$. The dependence of the charge density on $x_1$ for these three values of $x_3$ is illustrated in Fig.~\ref{fig:transverse_profiles}. The caloron symmetry axis goes through the centers of all the plots. Again, the coordinates were shifted and rescaled in order to match the caloron axis at $x_1 = L/2$, $x_2 = L/2$ on different lattices. The magnetic field strengths are the same as for the left plot of Fig.~\ref{fig:charge_densities_z_vs_size}.

 For all three values of $x_3$, the charge density is peaked around the caloron symmetry axis. There is also a large excess of positive imaginary charge density close to the boundaries of the lattice, which clearly diminishes as the lattice size increases. A comparison of the profiles for the $16^3 \times 4$ lattice and the rescaled $24^3 \times 6$ lattice shows that the profiles practically do not depend on the lattice spacing. This suggests that these boundary peaks are finite-volume, rather than discretization, artifacts. For asymptotically large volumes one can therefore expect that the charge density is localized near the caloron axis only.

 These findings can be readily interpreted along the lines of the previous Subsection \ref{subsec:const_field_qcd}. Namely, the monopole and the anti-monopole which constitute the caloron create a chromomagnetic field between them. In the vicinity of a mid-point this field is approximately constant, uniform, parallel to the caloron axis and proportional to the $\tau_3$ matrix in the color space. The nontrivial holonomy of the caloron solution, i.e.\ the Polyakov loop outside of the monopole and the anti-monopole, is also proportional to $\tau_3$ and thus plays the role of an imaginary isospin chemical potential. If a constant Abelian magnetic field parallel to the caloron symmetry axis is superimposed on such a field configuration, an imaginary electric charge should appear between the monopole and the anti-monopole.

 The origin of the charge excess which is localized directly on the monopole or the anti-monopole is not so clear to us. Possible explanation could be the following: since the monopole and the anti-monopole also create chromoelectric fields $\vec{E} \sim \pm \vec{r}/r^3 \, \tau^3$, in a state of thermal equilibrium the fermions would tend to decrease this electric field by forming Debye screening clouds of negative and positive $\tau_3$-charge, respectively. This screening clouds should be seen in the isospin density $\vev{\bar\psi\lr{x} \gamma_0\tau^3 \psi\lr{x}}$ (note that this is written in the particular gauge where the holonomy is in the $\tau^3$-direction). However, they would not contribute to the electric charge density as there would be an equal amount of particle as well as anti-particle states due to the $\tau_3$ symmetry. This argument, however, is no longer valid when the magnetic field is in place, as we have seen that the combined effect of magnetic field and nontrivial holonomy is to lift this symmetry. This would cause otherwise invisible Debye clouds to be seen in the charge density as bumps of positive/negative charge around the cores of the monopoles.

\subsection{An estimate of possible experimental importance}
\label{subsec:experiment}

 Let us now roughly estimate the importance of the described phenomenon for heavy-ion collisions. As the simplest approximation, we first adopt the picture of ``spaghetti'' QCD vacuum \cite{Nielsen:79:1}, assuming that it consists of domains with roughly constant chromomagnetic field $F$. According to (\ref{ChargeFluctUniformFinal}), the relative decrease of the dispersion of charge fluctuations is given by the ratio $\frac{\mymin{F^2, B^2}}{\mymax{F^2, B^2}}$. We estimate the typical strength of chromomagnetic field in QCD at temperatures close to the deconfinement phase transition as $F \sim 0.1 \, \GeV^2$ basing on the lattice measurements of the finite-temperature gluon condensate \cite{Miller:06:1}. As a moderate estimate of the magnetic field strength, we take $\lr{2 e/3} B = m_{\pi}^2 \approx 0.02 \, \GeV^2$ (here $2 e/3$ is the physical charge of the $u$ quark) which corresponds to Au-Au collision with impact parameter $b = 4 \, \fm$ at $\sqrt{s} = 200 \, \GeV$ \cite{Skokov:09:1, Skokov:13:1}. Then from (\ref{ChargeFluctUniformFinal}) it follows that the effect of magnetic field of such strength is to suppress the charge fluctuations by several percent. This estimate of course does not take into account any inhomogeneities of the magnetic and the chromomagnetic fields as well as different orientations of the latter in the color space and thus can only be considered as a qualitative prediction. Also higher magnetic fields will lead to much stronger suppression of charge fluctuations - according to (\ref{ChargeFluctUniformFinal}), the most significant suppression occurs when the strengths of external magnetic field and of chromomagnetic field are nearly equal. The effect might be also somewhat larger for real QCD due to larger degeneracy of states for $SU\lr{3}$ gauge group as compared to $SU\lr{2}$ gauge group considered here.

 As a more advanced model of QCD vacuum, one can consider a model of a dilute caloron gas \cite{Ilgenfritz:07:1}. We first fix the lattice spacing $a$ in physical units by using the expression for the distance between the monopole and the anti-monopole in the caloron $d = \pi \rho^2 T$, where $T$ is the temperature and $\rho$ is the caloron size parameter, for which we take a characteristic value $\rho = 0.33 \, \fm$ from \cite{Ilgenfritz:07:1}. Taking into account that for the $16^3 \times 4$, $20^3 \times 4$ and $24^3 \times 4$ lattices the distance between the monopole and the anti-monopole is $8$ lattice spacings, we obtain $a \approx 0.10 \, \fm$ for these lattices. Then from the linear fit on Fig.~\ref{fig:caloron_total_charge} we roughly estimate the total charge induced on a single caloron as $\Im J_0\lr{B} \approx c \, B$, where $B$ is the projection of the magnetic field on the caloron symmetry axis and $c \approx 2 \, \GeV^{-2}$. For a dilute gas of calorons and anti-calorons with concentration $n \approx 1 \, \fm^{-4}$ \cite{Ilgenfritz:07:1} in a four-volume $V_{4d}$ we estimate the contribution of the induced imaginary charge $\Im J_0$ to the expectation value of the squared total charge $Q$ as
\begin{eqnarray}
\label{ChargeFluctuations}
 \delta Q^2 \approx -1/3 \, \vev{\vev{ J_0^2\lr{B }}} \, n V_{4d}  ,
\end{eqnarray}
where $\vev{\vev{\ldots}}$ is the average over the caloron ensemble. Here we have assumed that the topological charges of all ``quasi-particles'' in a gas are statistically independent and take the values $\pm 1$ (corresponding to either a caloron or an anti-caloron) with equal probabilities. The factor
$$
 1/3 = \frac{1}{4 \pi} \int\limits_{-\pi}^{\pi} 2 \pi \, d \cos{\theta} \, \cos^2{\theta}
$$
comes from averaging over all possible caloron orientations in the three-dimensional space. For our simple estimate, we take $V_{4d}$ as the product of a typical three-dimensional volume of the fireball created in Au-Au heavy-ion collisions, $V_{3d} \sim \lr{5\, \fm}^3$, times the inverse critical temperature of the deconfinement phase transition $T_c \approx 200 \, \MeV \approx \lr{1 \fm}^{-1}$. The magnetic field is again estimated as $\lr{2 e/3} B \approx 0.02 \, \GeV^2$. With such collision parameters, we estimate the contribution of an imaginary charge induced on calorons to the fluctuations of the total charge as $\delta Q^2 \approx -0.05$. It can be easily changed by an order of magnitude, for example, if one takes into account the non-Poisson distribution of the number of calorons in a fixed volume \cite{Ilgenfritz:07:1}.

 We thus conclude that within the simple models which we have considered the effect of the magnetic field would be to decrease charge fluctuations in off-central heavy-ion collisions by several percent. Clearly, for off-central collisions it is practically impossible to change the value of the induced magnetic field without changing the volume of the fireball. As the impact parameter of the collision grows, the magnetic field induced by spectators also grows \cite{Skokov:09:1, Skokov:13:1}, but, on the other hand, the volume $V_{3d}$ of the fireball decreases. Thus one can expect that the few-percent-level effect of the decrease of charge fluctuations due to increasing magnetic field would not be visible behind the much stronger effect of decreasing charge multiplicities due to decreasing fireball volume $V_{3d}$. Therefore if it is possible at all to detect the described effect, it should primarily manifest itself in the centrality dependence of the volume-independent observables such as $D = \vev{Q^2}/\vev{N_{tot}}$ \cite{Jeon:00:1}, where $N_{tot} \sim V_{3d}$ is the total number of detected charged particles.

 We should also make a common cautionary remark concerning the use of the grand canonical ensemble approximation to describe the fluctuations of charge in heavy ion collisions. Namely, the total charge of all produced particles is of course conserved, but in experiment charged particles are only detected in a certain rapidity range. The charge of detected particles does fluctuate and for some reasonable choice of rapidity range these fluctuations can be described by a grand canonical ensemble of quark-gluon plasma at some finite chemical potential.

\section{Conclusions}
\label{sec:conclusions}

 In this paper we have described a mechanism by which a combination of magnetic field and isospin magnetic field can lead to the emergence of a net electric charge in a gas of Dirac fermions in the presence of isospin chemical potential or a non-trivial holonomy of a non-Abelian gauge field. If the system is described as having a constant chemical potential, that is, in terms of a grand canonical ensemble, the generated charge is borrowed from the environment to which the system is coupled and which keeps the chemical potential constant. We have also demonstrated that in case when no charge is exchanged with the environment, the charge density still appears due to spatial charge separation, so that the region with nonzero field strength acquires the excess of charge of one sign, and its exterior acquires the charge of opposite sign. In its simplest form the effect can be realized and observed in strained monolayer graphene, where a combination of mechanical strain, magnetic field and valley imbalance should lead to the emergence of nonzero electric charge.

 A more subtle case is that of non-Abelian gauge theory, in which the role of the isospin quantum number is played by the color of quarks and the nontrivial holonomy of gauge field configurations (Polyakov loop) is equivalent to an imaginary isospin chemical potential. In this case the induced charge is also imaginary for a fixed field configuration and vanishes after averaging over field configurations with all possible values of the holonomy. However, this imaginary charge still yields a negative contribution to the expectation value of the squared total charge in the system. We have explicitly studied the induced imaginary charge and its effect on charge fluctuations in the backgrounds of a constant chromomagnetic field and an $SU\lr{2}$ caloron with superimposed Abelian magnetic field. We found that the induced charge is localized between the monopole and the anti-monopole which constitute the caloron. Simple estimates made in Subsection \ref{subsec:experiment} suggest that the described effect should tend to slightly suppress charge fluctuations in off-central heavy-ion collisions. It might be also interesting to test the dependence of charge fluctuations on the external magnetic field in lattice QCD simulations. In this case one can measure the charge fluctuations with a very good precision, and the magnetic field strength can take quite high values \cite{Buividovich:09:7}.

\begin{acknowledgments}
 The authors are grateful to G.~Ba\c{s}ar, G.~Dunne, G.~Endr\H{o}di, M.~Katsnelson, A.~Sch\"{a}fer, E.~Shuryak and M.~Zubkov for valuable discussions and to the ECT$^*$ in Trento for the hospitality during the workshop ``QCD in strong magnetic fields''. The work of P.B. and F.B. was supported by the S.~Kowalewskaja award from the Alexander von Humboldt foundation (sponsored by the Ministry for Education and Research of the German Federal Republic) and by DFG (grants BR 2872/4-2 and BR 2872/5-1). T.S. is supported by BayEFG.
\end{acknowledgments}


\end{document}